\pdfoutput=1

\documentclass[
preprint,
amsmath,amssymb,
aps,
pre
]{revtex4-2}

\usepackage{derivative}
\usepackage{subcaption}
\usepackage{siunitx}
\usepackage{tikz}
\usepackage{graphicx}

\newcommand{\Psicr}{\Psi_{\mathrm{cr}}}
\newcommand{\psicr}{\psi_{\mathrm{cr}}}
\newcommand{\Psiel}{\Psi_{\mathrm{el}}}
\newcommand{\psiel}{\psi_{\mathrm{el}}}
\newcommand{\Psisb}{\Psi_{\mathrm{sb}}}
\newcommand{\psisb}{\psi_{\mathrm{sb}}}
\newcommand{\Psitot}{\Psi_{\mathrm{tot}}}
\newcommand{\ave}[2][]{\left\langle #2 \right\rangle_{#1}}
\newcommand{\norm}[1]{\left\lvert #1 \right\rvert}
\newcommand{\dnorm}[1]{\left\lVert #1 \right\rVert}
\newcommand{\mbepsilon}{\boldsymbol{\varepsilon}}

\newcommand{\mbe}{\boldsymbol{e}}
\newcommand{\mbq}{\boldsymbol{q}}
\newcommand{\mbt}{\boldsymbol{t}}
\newcommand{\mbu}{\boldsymbol{u}}

\newcommand{\mbx}{\boldsymbol{x}}
\newcommand{\mbI}{\boldsymbol{I}}
\newcommand{\mbK}{\boldsymbol{K}}
\newcommand{\mbR}{\boldsymbol{R}}
\newcommand{\mbW}{\boldsymbol{W}}
\newcommand{\cS}{\mathcal{S}}
\newcommand{\mnab}{\boldsymbol{\nabla}}
\newcommand{\transpose}{\mathsf{T}}
\DeclareMathOperator{\tr}{Tr}

\usepackage[unicode,bookmarks,colorlinks=true,%
citecolor=brown,%
urlcolor=olive,%
linkcolor=violet,%
plainpages=false,pdfpagelabels,%
bookmarksopen=true,pdfstartview={FitH},%
pdftitle={A Phase-Field Study of Desiccation Crack Pattern Maturation
under Drying--Wetting Cycles},pdfauthor={Hiroki Fukushima and Satoshi Yukawa}]{hyperref}

\begin{document}
\title{A Phase-Field Study of Desiccation Crack Pattern Maturation
under Drying--Wetting Cycles}
\author{Hiroki Fukushima and Satoshi Yukawa}
\affiliation{%
Department of Earth and Space Science, Graduate School of Science,
The University of Osaka, Toyonaka, Osaka 560-0043, Japan
}
\date{\today}
\begin{abstract}
The characteristic intersection angle of the desiccation crack relaxes from near \ang{90} toward \ang{120} under repeated drying--wetting cycles.
However, the theoretical understanding of this relaxation is insufficient, especially
the modeling of the drying--wetting cycles. Here we introduce a phase-field model of desiccation fracture, extending the model proposed in previous studies by adding crack healing and a scar effect left by past cracks.
By repeating drying--wetting cycles in a finite element simulation, we find that the angle distribution develops a growing peak near \ang{120} as the cycle number increases, consistent with experiments.
The standard deviation of the intersection angle from \ang{120} relaxes exponentially with a characteristic time of about 4 cycles. These results are consistent with experiments, although the uncertainty of the estimated characteristic time is large.
Crack energy dominates the total energy and also relaxes exponentially with nearly the same characteristic cycle as the angle relaxation. This decay is driven mainly by a shortening of the effective crack length rather than a change in effective fracture toughness.

\end{abstract}
\maketitle


\section{Introduction}

A wide variety of fracture patterns arise in nature from material shrinkage during drying or cooling. Crack patterns produced by such fracture processes have long attracted interest in nonequilibrium physics, and researchers have investigated their physical and statistical properties.
For example, studies of the cracks that form when a two-dimensionally spread paste of powder and water dries \cite{GND+2015} have revealed various statistical properties of the resulting crack patterns and their geometric structure \cite{CH1964, GK1994, AL1995, SdBGM2000, BPC2005, PWPO2009}, as well as memory effects in which stimuli applied to the paste before drying are reflected in the post-drying crack pattern \cite{NM2005, NM2006, NM2006a}.
As a related phenomenon, researchers have also studied the statistical properties of the three-dimensional structure of geological objects known as columnar joints, which form when three-dimensionally extended lava cools, solidifies, and undergoes volumetric shrinkage \cite{Mul1998a, Mul1998, TM2004a, GML2006}.

In both cases, the geometric shape of the cracks has been of particular interest. For desiccation fractures in two-dimensional pastes, researchers have extensively studied the distribution of crack intersection angles.
Experimentally, this angle distribution is governed by the thickness of the paste: thicker pastes show a distribution peaked near \ang{90}, whereas thinner pastes show more crack intersections near \ang{120} \cite{GK1994}.
This behavior originates from differences in the mode of elastic-energy release associated with differences in crack-generation dynamics: cracks that form sequentially tend to intersect at \ang{90} \cite{Wei1999}, whereas cracks that form simultaneously tend to intersect at \ang{120} \cite{GMM2009, GCA+2010, Goe2013}.  
The intersection angle also relaxes toward \ang{120} when once-opened cracks heal during a wetting process through repeated drying--wetting cycles.
In this setting, cracks that initially intersect at \ang{90} have been observed to relax toward \ang{120} intersections over repeated wetting--drying cycles, through processes such as the migration of crack junctions and the reordering of crack-generation sequence.

In this paper, we use computational simulation to investigate the experimentally observed relaxation of the crack intersection angle from \ang{90} to \ang{120} under repeated drying and wetting, with the aims of reproducing this behavior numerically and of gaining a deeper understanding of it from an energetic point of view.
Several numerical modeling studies of the desiccation fracture of pastes have been reported. These can be classified into three types, that is, continuum descriptions of the paste as an elastic body from a microscopic standpoint \cite{Kit1999, Jag2004, IY2014, HGD2020}, mesoscale descriptions based on spring-network-type models \cite{HSB1996, HNKK2017, HMTD2023, SNKK2025}, and macroscopic models based on stochastic descriptions \cite{IY2014a}.
Each approach has its own strengths and weaknesses, and the appropriate model should be chosen based on the phenomenon one wishes to understand.

To examine the angle relaxation under drying--wetting cycles, an energy-based study is required, as suggested by experiments; for this, modeling based on an elastic body is essential. 
As an early study based on elastic bodies related to the maturation of crack patterns,
Jagla reproduced the relaxation of crack junctions from T-shaped to Y-shaped configurations 
by a model based on a two-dimensional elastic body \cite{Jag2004}.
Hofmann \textit{et al.} investigated how initially right-angled crack junctions relax toward \ang{120}, using a three-dimensional elastic model \cite{HAB+2015}.
These theoretical studies of angle relaxation, however, are not under drying--wetting cycles.
In Jagla's work, relaxation occurs within a single drying process, and Hofmann's work considers the change in angle as the crack propagates in the depth direction. 
Haque \textit{et al.} \cite{HMTD2023}, by contrast, modeled drying--wetting cycles and reproduced the relaxation of the intersection-angle distribution under such cycling. However, they represented the paste using a mesoscale spring-network model. Therefore, the energetic analysis is insufficient.

In light of this, we introduce a shrinkage-fracture model that combines elasticity theory with a phase-field model, based on \cite{HGD2020}, and additionally model the drying--wetting cycle grounded in experimental observations.
Here, we describe the paste as a two-dimensional isotropic elastic body, and represent cracks as a damage field via the phase-field method, grounding the energetics in elasticity theory.
We further model the drying--wetting cycle based on experimental facts, and by repeating the cycle, we reproduce the angle relaxation, quantitatively evaluate the angle relaxation and angle distribution, and assess the process from an energetic point of view.

The remainder of this paper is organized as follows. Sections~\ref{sec:model} and~\ref{sec:numericalcalc} describe the model and numerical method used in this study and present the way in which the drying--wetting cycle and the effect of past cracks as scars are incorporated into the continuum model.
Section~\ref{sec:results} provides a quantitative evaluation of the angle-relaxation process and the energy relaxation based on the numerical results.
Section~\ref{sec:summary} summarizes the findings of this study and discusses their implications.


\section{Model}
\label{sec:model}

In this section, we describe the phase-field-based desiccation-fracture crack model used in this study from the standpoints of the energy functional, the desiccation process, the irreversibility of crack generation, material heterogeneity, crack healing during wetting, and the effects of scars generated during the previous drying process.
Except for modeling crack healing during wetting and the effect of scars, the desiccation-fracture model is equivalent to that introduced in the previous study \cite{HGD2020}.

\subsection{Energy Functional}
\label{sec:energyfunctional}

We consider a thin paste, a mixture of powder and water, resting on a rigid substrate. We denote the region occupied by this paste as $\Omega$ with thickness $h$ and area $A$.
We align the $z$-axis with the thickness direction and the horizontal $xy$-plane with the paste plane. The $xy$-plane of the region $\Omega$ is taken to be a square of side length $L$, with periodic boundary conditions. As the system energy, we consider the elastic energy of the paste, the crack energy, and the interaction energy with the substrate.

First, consider the elastic energy of the paste. We assume the paste is sufficiently thin and uniform in the thickness direction and can be described as a two-dimensional linear isotropic elastic body. Its Lam\'e constants $\lambda,\mu$, together with the Young's modulus $E$, Poisson's ratio $\nu$, and the 2D bulk modulus $K$, satisfy the relations
\begin{equation}
\lambda = \frac{E\,\nu}{1-\nu^2},\qquad
\mu = \frac{E}{2(1+\nu)},\qquad
K = \lambda+\mu.
\end{equation}

At the in-plane horizontal coordinates $\mbx=(x,y)$ within this region, we define the displacement field $\mbu(\mbx)=(u_{x}(\mbx),u_{y}(\mbx))$, a two-dimensional vector field associated with the elastic deformation, and the damage field $d(\mbx) \in [0,1]$, a scalar field representing the crack.
The damage field is such that $d=0$ corresponds to a state with no crack whatsoever, and $d=1$ to a state in which a crack has formed and the material is sufficiently fractured.
The strain tensor $\mbepsilon$ associated with the displacement field is
\begin{equation}
\mbepsilon(\mbu)= \frac{1}{2}\left(\nabla\boldsymbol{u}+(\nabla\mbu)^{\transpose}\right)
=
\begin{pmatrix}
\varepsilon_{xx} & \varepsilon_{xy}\\
\varepsilon_{yx} & \varepsilon_{yy}
\end{pmatrix}.
\end{equation}
The individual components of the tensor are $\varepsilon_{xx} = \pdv{u_{x}}!{x},$
$\varepsilon_{yy}=\pdv{u_{y}}!{y}$
and $\varepsilon_{xy} = \varepsilon_{yx} = \frac{1}{2} \left(\pdv{u_{y}}!{x} + \pdv{u_{x}}!{y} \right)$.

We express the elastic energy in the region $\Omega$ as a volume integral of an elastic-energy density that is a functional of $\mbu$ and $d$. Due to the presence of the crack, the elastic-energy density is modified from the energy density ordinarily used for elastic bodies.
Here we adopt the expressions of \cite{AMM2009} and \cite{MHW2010}.
We first define the bracket notation $\langle X \rangle_{\pm}$ for a scalar quantity $X$ as
\begin{equation}
  \langle X\rangle_{\pm}
  \equiv \dfrac{X\pm\norm{X}}{2}
  = \begin{cases}
    \max (X,0) & \text{for $+$,}\\
    \min (X,0) & \text{for $-$.}
  \end{cases}
\end{equation}
This bracket notation is mutually complementary, and by using it, the volumetric strain can be expressed as
\begin{equation}
\tr \mbepsilon = \langle \tr \mbepsilon \rangle_{+} + \langle \tr \mbepsilon \rangle_{-}.
\end{equation}
Here, if $\tr \mbepsilon >0$ then $\tr \mbepsilon = \langle \tr \mbepsilon \rangle_{+}$, and if $\tr \mbepsilon <0$ then $\tr \mbepsilon = \langle \tr \mbepsilon \rangle_{-}$; this bracket notation thus allows the expansive and contractive components of the volumetric deformation to be extracted. We define the shearing strain $\mbe$ as
\begin{equation}
\mbe \equiv  \mbepsilon - \dfrac{1}{2} (\tr \mbepsilon) \mbI
=
\begin{pmatrix}
  \dfrac{\varepsilon_{xx}-\varepsilon_{yy}}{2} & \varepsilon_{xy}\\
  \varepsilon_{xy} &  \dfrac{-\varepsilon_{xx}+\varepsilon_{yy}}{2}
\end{pmatrix},
\end{equation}
where $\mbI$ is the identity tensor. With this, the elastic-energy density can be expressed, for a general isotropic elastic body, as the sum of the two terms
\begin{align}
\psiel^{+}(\mbepsilon) & \equiv \dfrac{K}{2} \ave[+]{\tr(\mbepsilon)}^{2}
+ \mu \mbe : \mbe,\\
\psiel^{-}(\mbepsilon) & \equiv \dfrac{K}{2} \ave[-]{\tr(\mbepsilon)}^{2}.
\end{align}

Here $\psiel^+$ represents the energy density that promotes crack opening and sliding, combining the volumetric expansion and shear that tend to open the crack surfaces.
In contrast, $\psiel^-$ represents the compressive energy density associated with the crack surfaces being pressed together.
Cracks readily open under tension or shear, whereas under compression they close and transmit compressive force in the normal direction. To express this physical feature in terms of energy, we model the system such that only the shear-energy component arising from positive volumetric deformation and the shearing strain is weakened by the presence of the crack \cite{HGD2020}, while the energy from the compressive volumetric component is left unaffected by the crack.
Introducing a degradation function $g(d)$ \cite{AT1990, BFM2000, AMM2009, MHW2010} that represents this weakening as a function of the damage field $d(\mbx) \in[0,1]$, we write the elastic-energy density as
\begin{equation}
\psiel(\mbepsilon(\mbu),d)=
g(d)\psiel^+(\mbepsilon)  +\psiel^-(\mbepsilon).
\label{eq:elasticenergydensity}
\end{equation}
Here, the degradation function $g(d)$ satisfies $g(0)=1$ for the case where no crack whatsoever has formed, and $g(1)=\eta, \enspace (0 < \eta \ll 1)$ for the case where the material is completely fractured. $\eta$ is a sufficiently small numerical parameter that prevents numerical instability. 
In addition, $g(d)$ is a monotonically decreasing function satisfying $g'(d)\le0$. Details of this degradation function are given in App.~\ref{sec:degradationFunc}.

Next, we model the crack energy following Griffith \cite{Gri1921}. Generating a new crack surface $\Gamma$ requires an energy proportional to the area of that crack surface; writing the fracture toughness of the material as $G_c$, this energy can be written as
\begin{equation}
    \Psicr^{\mathrm{Gr}}[\Gamma]=\int_{\Gamma}G_c \odif{A}.
\end{equation}
Since this expression requires explicit treatment of the crack surface $\Gamma$, it is difficult to track crack nucleation and complex propagation in a numerical analysis. Therefore, the phase-field method expresses the crack energy using the damage field $d$ as
\begin{equation}
\Psicr[d]
=\int_{\Omega}\psicr(\nabla d,d)\odif{V},
\label{eq:crack-energy-functional}
\end{equation}
where
\begin{equation}
\psicr
=\dfrac{G_c}{c_0\ell_0}\left(w(d)+\ell_{0}^2 \dnorm{\nabla d}^2\right)
\end{equation}
and $\ell_0$ is a regularization length that relates to the characteristic width of the crack, and $c_0$ is a constant chosen so that, in the limit $\ell_0\to0$,
\begin{equation}
     \Psicr[d] \to \Psicr^{\mathrm{Gr}}[\Gamma]
\end{equation}
holds. In this model, following the previous studies \cite{PAMM2011,BMMS2014,HGD2020}, we use $w(d)=d$ and $c_0=8\slash 3$.

Next, we consider the energy arising from the interaction between the paste and the substrate. In desiccation-fracture experiments on coffee paste \cite{GK1994}, the main contribution to the stress required for crack generation is the frictional force arising between the paste and the substrate.
In this model, we approximate the frictional force from the substrate as proportional to the horizontal displacement, assuming the paste is thin. Writing the proportionality constant as $\kappa$, the frictional force can then be written as the shear stress
\begin{equation}
  \mbt_{\mathrm{sb}} = g(d) \kappa \mbu.
\end{equation}
Here we use the same degradation function $g(d)$ as in the elastic energy, reflecting that where the crack has progressed, the substrate force is simultaneously reduced.
Using the paste thickness $h$, the interaction energy with the substrate can then be written as
\begin{equation}
  \Psisb[\mbu,d]
= \int_{\Omega} \psisb(\mbu,d)\odif{V},
  \label{eq:foundation_energy}
\end{equation}
where
\begin{equation}
  \psisb(\mbu,d) = \dfrac{\kappa}{2h} g(d)  \dnorm{\mbu}^{2}.
\end{equation}
Here the paste thickness $h$ is assumed to be sufficiently small compared with the other elastic length scales. $\kappa h$ has the dimension of a shear modulus.
When the three-dimensional elastic body of the paste is approximated as sufficiently thin, $\kappa h$ takes a value comparable to the shear modulus of the paste \cite{IY2014}.
Alternatively, within the shear-lag model \cite{LHPS2003}, it can be expressed using the shear modulus of the elastic foundation under the paste.
Here, we take $\kappa h$ to be comparable to the paste's shear modulus.

\subsection{Desiccation Process}
\label{sec:desiccation-process}

Next, we model the volumetric shrinkage accompanying paste desiccation. Since we assume a two-dimensional elastic body, we express the volumetric shrinkage by an in-plane isotropic strain tensor,
\begin{equation}
\mbepsilon^{\mathrm{sh}} \equiv \varepsilon_{\mathrm{sh}} \mbI,
\end{equation}
where $\varepsilon_{\mathrm{sh}}$ is a scalar quantity and $\mbI$ denotes the identity tensor. When $\varepsilon_{\mathrm{sh}}<0$, this represents the tendency of the paste to shrink isotropically within the plane.

To incorporate the effect of this shrinkage into the energy, the strain tensor appearing in the elastic-energy density \eqref{eq:elasticenergydensity} defined in the previous subsection must be replaced by the effective strain tensor obtained by subtracting the shrinkage contribution from the strain tensor computed directly from the displacement,
\begin{equation}
  \widetilde{\mbepsilon}(\mbu)
  = \mbepsilon(\mbu) - \mbepsilon^{\mathrm{sh}} = \mbepsilon(\mbu)- \varepsilon_\text{sh} \mbI.
\end{equation}

\subsection{Prohibition of Crack Healing in the Desiccation Process}
\label{sec:prohibit-crack-healing}

The generation and propagation of the crack are determined by minimizing the energy functional. However, if the energy functional is minimized naively, the damage field $d$ may decrease again.
This would phenomenologically represent a once-opened crack closing again and the material recovering as an elastic body, which is not appropriate for crack generation during desiccation.

Therefore, following the phase-field model of Miehe et al. \cite{MHW2010}, we introduce a history field to determine the damage field $d$. It prevents crack healing during desiccation and ensures a monotonic increase in the damage field.
The history field $H(\mbx,t)$, defined at each point $\mbx$ in space, is taken to retain the maximum value of the tensile component of the elastic energy experienced at that point between the initial state $t=0$ and the current time $t$,
\begin{equation}
  H(\mbx,t)
  \equiv \max_{0\le \tau \le t}\,\psiel^{+}\left(
  \widetilde{\mbepsilon}(\mbu(\mbx,\tau))
  \right).
  \label{eq:history-strain-energy}
\end{equation}

Using this history field $H(\mbx,t)$, we modify a form of the elastic-energy density introduced in Subsec.~\ref{sec:energyfunctional} as
\begin{equation}
\psiel^{(d)}(\widetilde{\mbepsilon},d)\equiv
g(d)H(\mbx,t) + \psiel^{-}\left(\widetilde{\mbepsilon}(\mbu)\right).
\end{equation}
In this way, the damage field $d$ can no longer decrease during the energy-minimization process by using the history field $H(\mbx,t)$ in place of $\psiel^{+}(\widetilde{\mbepsilon})$. It represents the irreversibility of the crack generation.

The energy functional finally adopted in this model takes the following forms: Equation~\eqref{eq:total-energy-functional-for-u} is used to determine the displacement field $\mbu$, and Equation~\eqref{eq:total-energy-functional-for-d} is used to determine the damage field $d$:
\begin{equation}
\Psitot[\mbu,d] =
   \int_{\Omega} \psiel\left(\widetilde{\mbepsilon}(\mbu),d\right)\odif{V}
  + \int_{\Omega} \psicr(\nabla d,d)\odif{V}
  + \int_{\Omega} \psisb(\mbu,d)\odif{V},
  \label{eq:total-energy-functional-for-u}
  \end{equation}
\begin{equation}
\Psitot^{(d)}[\mbu,d] =
   \int_{\Omega} \psiel^{(d)}\left(\widetilde{\mbepsilon}(\mbu),d\right)\odif{V}
  + \int_{\Omega} \psicr(\nabla d,d)\odif{V}
  + \int_{\Omega} \psisb(\mbu,d)\odif{V}.
  \label{eq:total-energy-functional-for-d}
    \end{equation}
In this model, we compute the crack-propagation behavior of the shrinking paste by finding the displacement field $\mbu$ and the damage field $d$ that minimize these energy functionals.

\subsection{Randomness of the Paste}
\label{sec:initial-randomness}

Next, we formulate the heterogeneity within the paste.
Within the paste, due to differences in local texture and defect distribution, it is expected that the fracture toughness $G_c$ introduced in the crack energy Eq.~\eqref{eq:crack-energy-functional} and the critical energy density $\psi_c$ for crack nucleation, introduced in the degradation function $g(d)$ (see App.~\ref{sec:degradationFunc}, Eq.~\eqref{eq:g_def}), are not spatially uniform.
We also expect the fracture toughness and critical energy density to vary spatially with a finite correlation length and to be correlated. 
Hu et al. \cite{HGD2020} modeled this material heterogeneity using two mutually correlated random fields, $G_c(\mbx)$ and $\psi_c(\mbx)$, that are somewhat spatially smooth, and examined the effect of the strength of this correlation and of the smoothing on the crack pattern.
In the present model, we also model the paste heterogeneity as two mutually correlated smooth random fields.

\begin{figure}
  \begin{subfigure}[t]{0.49\linewidth}
    \centering
    \includegraphics[width=0.80\linewidth]{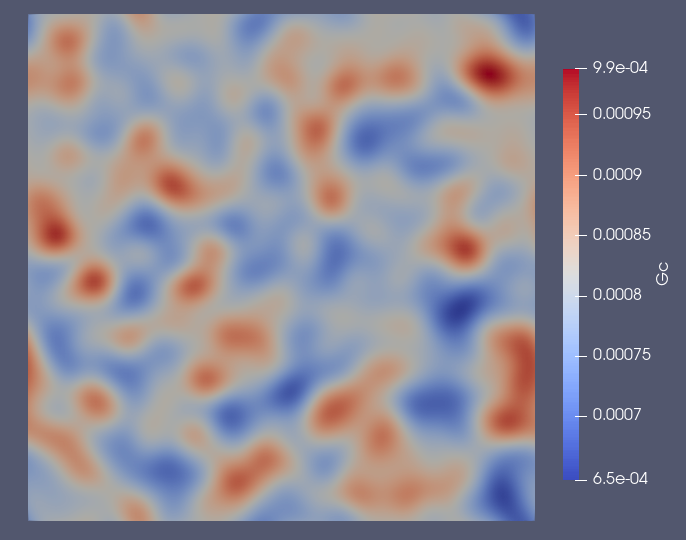}
    \caption{Spatial distribution of the fracture toughness $G_c$.}
    \label{Fig:Gc}
  \end{subfigure}\hfill
  \begin{subfigure}[t]{0.49\linewidth}
    \centering
    \includegraphics[width=0.81\linewidth]{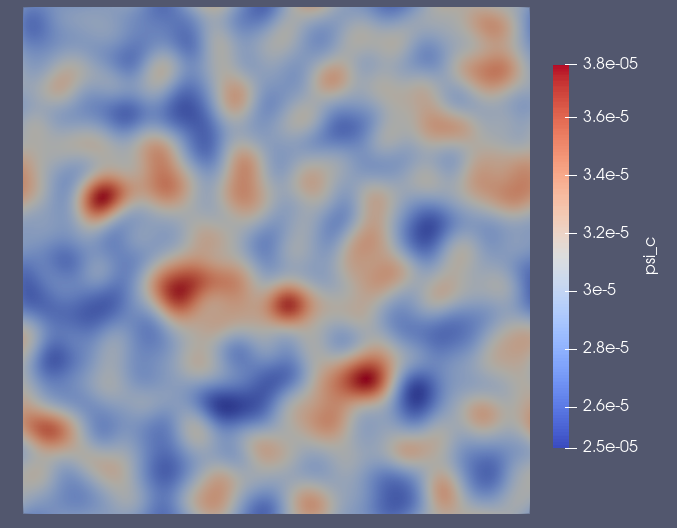}
    \caption{Spatial distribution of the critical energy density $\psi_c$.}
    \label{fig:Psi_c}
  \end{subfigure}
  \caption{(color online) Spatial distributions of the fracture toughness $G_c$ and the critical energy density $\psi_c$, visualized using the open-source visualization software ParaView \cite{ParaView}.}
  \label{fig:GcPsic}
\end{figure}
Details of the procedure are expressed in App.~\ref{sec:randomfield}.
Figure~\ref{fig:GcPsic} shows an example for the case of an initial correlation $\rho=0.5$ and $N_{s}=64$ smoothing repetitions (parameters are defined in App.~\ref{sec:randomfield}).
Although the two fields exhibit a similar global pattern, their resulting spatial distributions do not coincide locally.

\subsection{Modeling of Crack-induced Degradation and Healing after the Desiccation Process}

This study aims to examine the maturation of the crack pattern generated during the drying--wetting cycle.
Therefore, it is necessary to model both the paste's healing during wetting and the weakening effect that cracks generated during the drying process have on the paste in the next cycle.
According to Goehring et al. \cite{GCA+2010}, the scars left by the generated crack lines play an important role in crack formation in the next generation and also in the maturation of the crack pattern.
It has been reported that a shallow depression forms around the old crack when a previously formed crack pattern is rewetted and the crack closes. This shallow depression leaves a region where the material thickness is about $10\%$ thinner than the surroundings.
These scars act as weak lines that readily reopen in the next drying process, making it likely that a crack will again form along the same line in the next generation.
In this study, we formalize this experimental fact that a trace persisting even after closure induces recracking in the next cycle within the phase-field framework as an internal state updated between drying--wetting cycles.

Explicitly introducing a change in material thickness, or modeling the shallow depression itself, is difficult within the current two-dimensional continuum formalism.
We therefore newly introduce a scalar field representing the path of past cracks in the model:
\begin{equation}
    S(\mbx) \in [0,1],
\end{equation}
which we call the memory field. We take $S(\mbx)= 0$ to represent a location that no crack has previously passed through and at which essentially no scar remains, and $S(\mbx) = 1$ to represent a location that a crack has previously passed through and at which a scar remains. Using this scalar field, we express the degree to which the material has become slightly thinner and the local fracture resistance has decreased, not through the material thickness itself, but indirectly through a weakening of the fracture toughness $G_c(\mbx)$ and critical energy density $\psi_c(\mbx)$.

We set the memory field as follows. At the end of a given drying process, the damage field $d(\mbx) \in [0,1]$ has been computed. Here we use this damage field directly as the memory field for the next cycle, i.e., we set the memory field $S(\mbx)$ to
\begin{equation}
  S(\mbx) = d(\mbx).
  \label{eq:def_S_simple}
\end{equation}
We incorporate the weakening effect on the next drying process arising from the memory field $S(\mbx)$ by reducing the fracture toughness $G_c(\mbx)$ and critical energy density $\psi_c(\mbx)$ in proportion to $S(\mbx)$.
Specifically, these effective values $G_c^{\mathrm{eff}}(\mbx)$ and $\psi_c^{\mathrm{eff}}(\mbx)$ are taken to be
\begin{align}
G_c^{\mathrm{eff}}(\mbx) &= \left(1 - a_G S(\mbx)\right) G_c(\mbx),
\label{eq:effective_gc}\\
\psi_c^{\mathrm{eff}}(\mbx) &= \left(1 - a_P S(\mbx)\right) \psi_c(\mbx).
\end{align}
Here $a_G, a_P \in [0,1)$ are dimensionless parameters that respectively adjust the degree to which the fracture toughness and the critical energy density are reduced by the scar. 
In regions with $S(\mbx) = 0$, we have $G_c^{\mathrm{eff}}(\mbx) = G_c(\mbx)$ and $\psi_c^{\mathrm{eff}}(\mbx) = \psi_c(\mbx)$, so that these regions are unaffected by any existing crack; as $S(\mbx)$ increases, however, $G_c^{\mathrm{eff}}(\mbx)$ and $\psi_c^{\mathrm{eff}}(\mbx)$ decrease continuously, lowering the energy required for fracture to progress.
A smaller value of $G_c^{\mathrm{eff}}(\mbx)$ means that the energy cost of creating a new crack surface becomes smaller as the fracture toughness weakens, while a smaller value of $\psi_c^{\mathrm{eff}}(\mbx)$ means a reduction in the local stiffness of the material as an elastic body, mediated through the degradation function $g(d)$.

\section{Numerical Calculation: Cycle of Desiccation and Healing}
\label{sec:numericalcalc}

We now describe how to repeat the drying and wetting cycle in the simulation.
First, as the initial state, the random fields $G_{c}(\mbx), \psi_{c}(\mbx)$ are generated by the method described in App.~\ref{sec:randomfield}. We retain these random fields unchanged throughout the subsequent cycles.
Next, the displacement field $\mbu(\mbx)$, damage field $d(\mbx)$, and history field $H(\mbx)$ are set to $0$. We compute $\mbu(\mbx), d(\mbx)$, and $H(\mbx)$ that minimize the energy functional
while decreasing the volumetric shrinkage strain $\varepsilon_\mathrm{sh}$ as described in Sec.~\ref{sec:desiccation-process} for the drying process.
Once the target volumetric strain is reached, the first-generation drying process ends.
In the healing process, the memory field $S(\mbx)$ is constructed from the final damage field $d(\mbx)$, and $G_c^{\mathrm{eff}}(\mbx), \psi_c^{\mathrm{eff}}(\mbx)$ are then constructed from the retained random fields $G_{c}(\mbx), \psi_{c}(\mbx)$ based on $S(\mbx)$.
These $G_c^{\mathrm{eff}}(\mbx), \psi_c^{\mathrm{eff}}(\mbx)$ are used as the random fields in the drying process of the next generation.
The uniform shrinkage strain $\varepsilon_\mathrm{sh}$ is then reset to zero, and the displacement field $\mbu(\mbx)$, damage field $d(\mbx)$, and history field $H(\mbx)$ are again reset to $0$.
We treat a drying-and-healing set as a single cycle and repeat it to simulate the process.

In this way, we formulate the present study as a cycle in the phase-field shrinkage-fracture model.
We solve the phase-field model at each cycle and update the local material properties using the memory field $S$, constructed from the damage field at the end of the cycle.
This framework allows us to incorporate the effect whereby scars from the previous generation induce recracking in the next generation as an internal variable in the continuum fracture model.
We perform free-energy minimization during the drying process using the finite element method. Details of this numerical calculation are given in App.~\ref{sec:details-calc}. The parameters used in this study are summarized in Table~\ref{tb:numerical-parameters}.
\begin{table*}
	\centering
  \begin{ruledtabular}
  \begin{tabular}{llcl}
    Parameter & Symbol & Value & Unit \\
    \hline
    Young's modulus & $E$ & 4 & \unit{MPa} \\
    Poisson's ratio & $\nu$ & 0.2 & --  \\
    Regularization length & $\ell_0$ & 0.5 & \unit{mm} \\
    Maximum shrinkage strain & $\varepsilon_{\mathrm{sh}}^\text{max}$ & 0.05 & -- \\
    Thickness & $h$ & 4 & mm \\
    Proportionality constant of frictional force & $\kappa$ & 0.4 & \unit{N/mm^3}\\
    \hline
    Mean fracture toughness & $\overline{G_c}$ & $8\times10^{-4}$ & \unit{mJ/mm^2} \\
    Mean critical energy density & $\overline{\psi_c}$ & $3\times10^{-5}$ & \unit{mJ/mm^2} \\
    Relative standard deviation of fracture toughness&    $c_{G_c}^{(0)}$      &  1 &--  \\
		Relative standard deviation of critical energy density&    $c_{\psi_c}^{(0)}$   & 1  &--  \\
    Number of smoothing steps for random field & $N_s$ & 64 & --\\
    Correlation coefficient of random field & $\rho$ & 0.5 & --  \\
       \hline
    Weakening coefficient for fracture toughness & $a_G$ & 0.15 & -- \\
    Weakening coefficient for critical energy density & $a_P$ & 0.35 & --  \\
    \hline
    System size ($x$ direction) & $L_x$ & 100 & \unit{mm} \\
    System size ($y$ direction) & $L_y$ & 100 & \unit{mm} \\
    \hline
    Number of loading steps per cycle & $N_{\text{step}}$ & $1000$ & -- \\
    Shrinkage strain per step & $\Delta \varepsilon_{\text{sh}}=\varepsilon_\text{sh}^\text{max}/N_\text{step}$ & $5 \times 10^{-5}$ & -- \\
    Discretization Number ($x$ direction) & $N_{x}$ & 128 & --\\
    Discretization Number ($y$ direction) & $N_{y}$ & 128 & --
  \end{tabular}
\end{ruledtabular}
  \caption{Parameters of the phase-field shrinkage-fracture model\label{tb:numerical-parameters}}
\end{table*}


\section{Results}
\label{sec:results}
\subsection{Snapshots}

\begin{figure}
	\centering
	\includegraphics[width=0.8\linewidth]{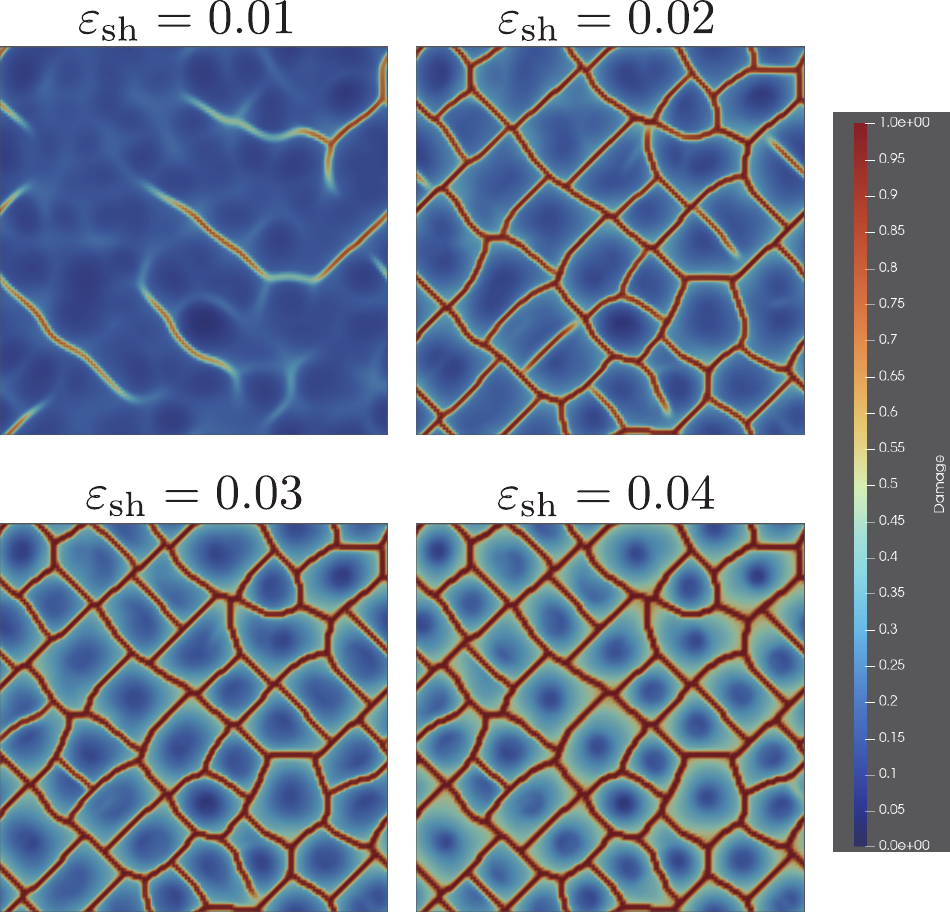}
	\caption{(color online) Snapshots of crack-pattern growth during the drying process of the first cycle. The damage field is shown as a color map, with $d\simeq 1$ representing a crack. Cracking begins at around $\varepsilon_\text{sh} \sim 0.01$. Subsequently, each fragment progressively subdivides.}
	\label{Fig:crack-snapshot}
\end{figure}
\begin{figure}
	\centering
  \includegraphics[width=\linewidth]{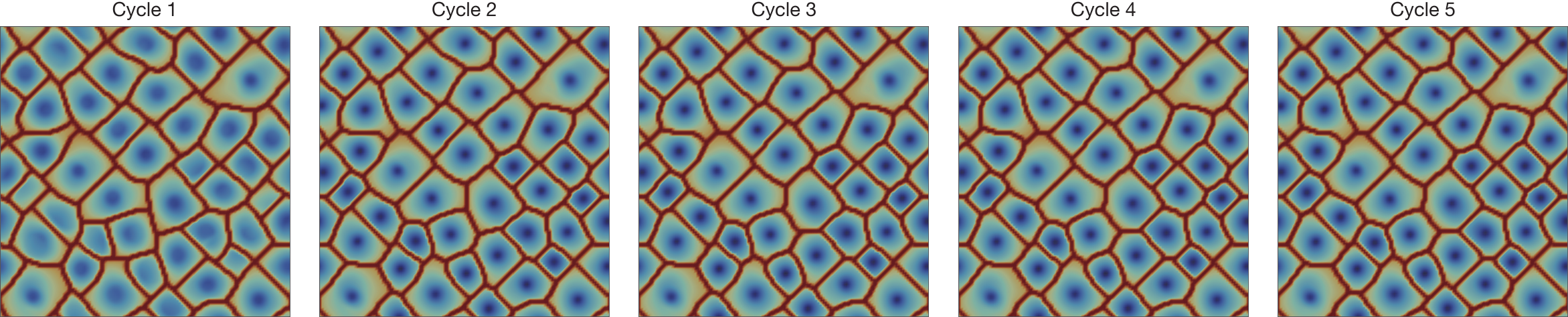}
	\caption{(color online) Snapshots of the damage field of the crack pattern at the end of each cycle. In the first cycle, cracks form sequentially while curving, and later cracks meet existing cracks at right angles. From the second cycle onward, cracks form simultaneously at multiple locations, and as the cycle proceeds, the previously curved cracks appear to run in straight lines.}
\label{Fig:crack}
\end{figure}
We first plot the crack generation during the initial drying process (the first cycle) as snapshots of the damage field for each value of the shrinkage strain in Fig.~\ref{Fig:crack-snapshot}. Cracking begins at a shrinkage strain of about $\varepsilon_\text{sh} \sim 0.01$. Subsequently, polygonal fragments are generated, and as the strain increases further, each fragment subdivides.

As described in Sec.~\ref{sec:model}, the damage field is copied to the memory field and healed, and the cycle of the drying process is repeated. Figure~\ref{Fig:crack} shows the damage field at the end of the drying process for each cycle. In the first cycle, cracks formed sequentially and curved, with later cracks meeting existing ones at right angles. From the second cycle onward, cracks more often arise simultaneously at multiple locations, and, as shown in the figure, the previously curved cracks run increasingly straight as the cycle proceeds. Because the strength is weakened at the locations where cracks were generated in the previous cycle, similar cracks are seen to be regenerated even after healing. However, the reproduced cracks are not identical; crack junctions migrate and rearrange, and cracks straighten.

\begin{figure}
  \centering
	\includegraphics[width=0.80\linewidth]{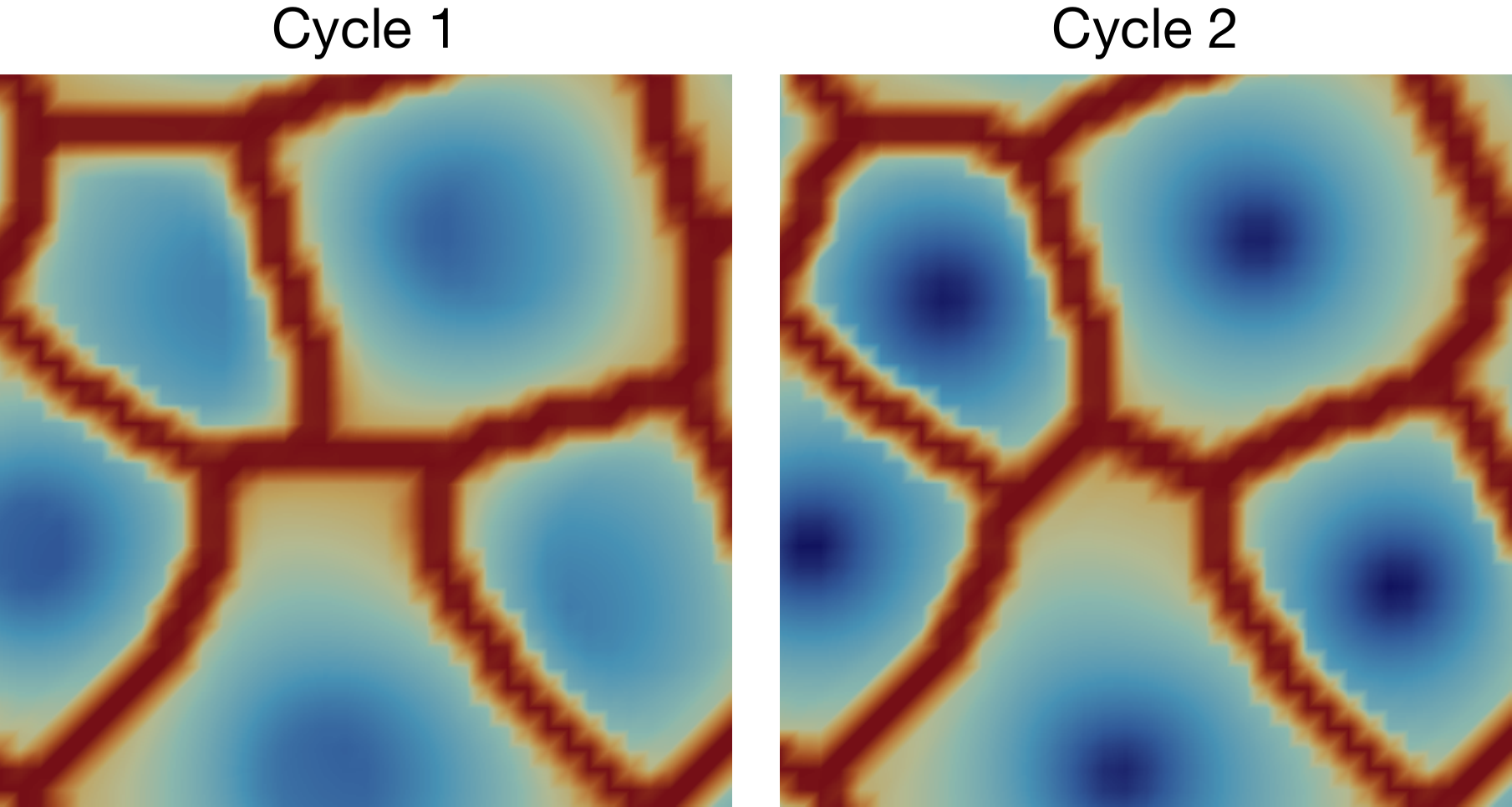}
  \caption{(color online) Magnified view of the same region at the end of the first and second cycles. In the first cycle, cracks form sequentially, and later cracks tend to meet existing cracks at right angles. In the second cycle and beyond, however, multiple cracks tend to form simultaneously, accompanied by a change in the crack intersection angle.}
  \label{fig:magcrack}
\end{figure}
As an example illustrating the change in shape of a crack junction, Figure~\ref{fig:magcrack} shows a magnified view of the same region at the end of the first and second cycles. In the first cycle, cracks formed sequentially, whereas from the second cycle onward multiple cracks tend to form simultaneously. This is accompanied by a change in the crack intersection angle, as shown in the figure. Thus, in this model, the crack pattern changes as the drying--wetting cycle is repeated. In the following subsections, we evaluate this behavior quantitatively.

\subsection{Statistics of Crack Angles}

The snapshots show that the pattern changes as the cycle repeats. To evaluate this change quantitatively, we numerically identify the crack junctions and measure the angles formed by the cracks. By carrying this out for the final state at each cycle, we examine the quantitative change in the angle distribution.
The results below are calculated from 20 samples.
We construct the angular distribution function $p_c(k)$ at cycle $c$ from the histogram $n_{s,c}(k)$ ($k$ is the angle bin) for sample $s$ at cycle $c$.
We compute the distribution function $p_{s,c}(k)$ for sample $s$ at cycle $c$ as
\begin{equation}
    p_{s,c}(k)=\frac{n_{s,c}(k)}{\sum_{k}n_{s,c}(k)}.
\end{equation}
Taking the sample average of this distribution, we get the angular distribution function as
\begin{equation}
p_{c}(k) = \ave{p_{s,c}(k)}.
\end{equation}
Appendix~\ref{sec:angle-algorithm} summarizes the method used to measure the angles.

\begin{figure}
    \centering
    \includegraphics[width=0.8\linewidth]{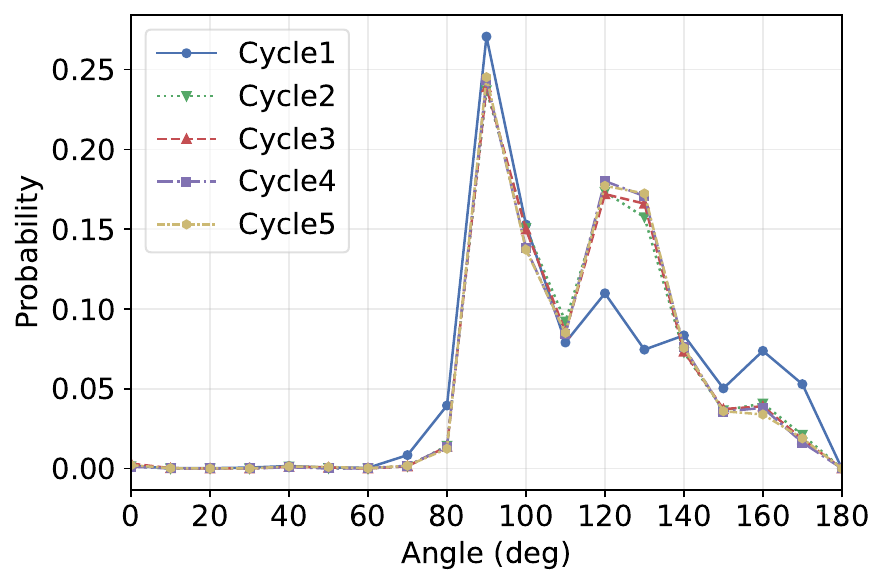}
    \caption{Sample average of the angle distributions for cycles 1 through 5. We normalized each sample's angle histogram by its total count and averaged them to obtain a probability distribution. Error bars are also plotted. In the first cycle, a pronounced peak appears near \ang{90}, while the frequency in the range \ang{110}--\ang{150} is low. As the cycle proceeds, however, the frequency near \ang{120} $\sim$ \ang{140} increases, while the \ang{90} peak persists.}
    \label{fig:angle-hist-cyc1-5}
\end{figure}
The change in the angle distribution from cycles 1 through 5 is shown in Fig.~\ref{fig:angle-hist-cyc1-5}.
Figure~\ref{fig:angle-hist-cyc1-5} shows that the angle distribution clearly differs between the first cycle and the subsequent cycles. 
In the first cycle, the peak near \ang{90} and the broad peak in the range \ang{160}$\sim$\ang{170} are observed, which means that T-shaped intersections predominate. In subsequent cycles, however, the distribution near \ang{120} $\sim$ \ang{140} increases, indicating an increase in Y-shaped intersections.

These results show that the frequency of angles near \ang{120} increases with repeated drying--wetting cycles. We also found that intersections near \ang{90} do not decrease very much in this model. Furthermore, examining the distribution up to 50 cycles revealed that the angle distribution converges almost completely by around cycle 5, after which it changes very little with further cycling.

\begin{figure}
  \centering
  \includegraphics[width=0.8\linewidth]{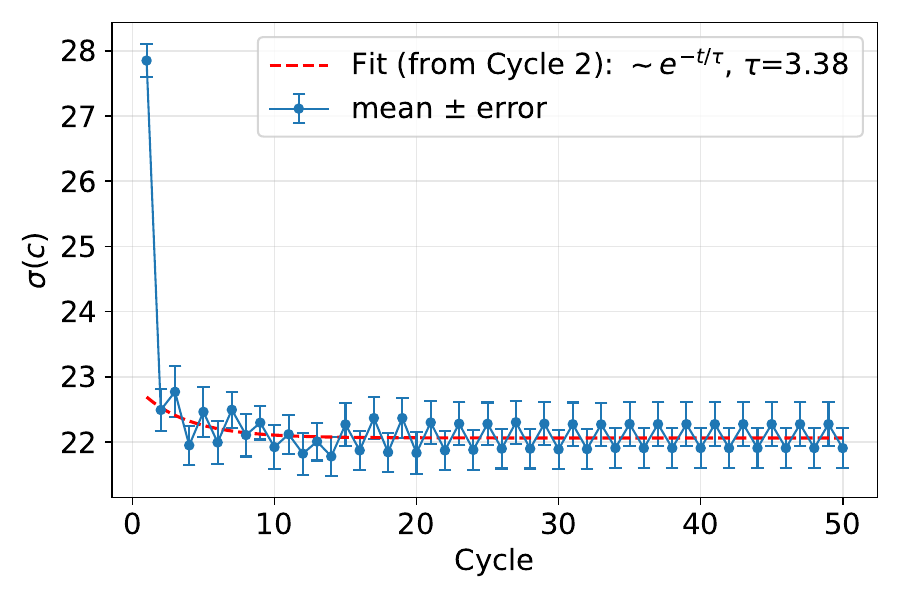}
  \caption{Time series of the sample-averaged standard deviation $\sigma$ from \ang{120}, together with an exponential fit. Fitting the time series to $\sigma(c) = \sigma_\infty + A_\sigma\exp(-c/\tau_\theta)$ for cycles $c\ge 2$ onward gave a relaxation time of $\tau_\theta=3.38\pm 3.4$.
  \label{fig:sigma_120_expfit}}
\end{figure}
For a further quantitative evaluation, following the study by Goehring et al. \cite{GCA+2010}, we compute the standard deviation from a Y-shaped intersection as the standard deviation of the angle $\theta$ from \ang{120}.
First, we compute
\begin{equation}
  \sigma^2_{s}(c) = \frac{1}{N(s,c)}\sum_{i=1}^{N(s,c)}
  \left(\theta_i-\ang{120}\right)^2
\end{equation}
for each cycle within each sample, where $N(s,c)=\sum_{k}n_{s,c}(k)$ is the number of angles for each sample and cycle.
Taking the square root of this and averaging over the 20 samples,
we get the standard deviation $\sigma(c)$ from a Y-shaped intersection
\begin{equation}
  \sigma(c) = \ave{\sqrt{\sigma^{2}_{s}(c)}}.
\end{equation}
We plot this quantity together with its error in Fig.~\ref{fig:sigma_120_expfit}.
This result shows that $\sigma$ decays substantially at first and subsequently settles to a constant value, indicating that the crack intersection angle is substantially reorganized early on and matures from T-shaped to Y-shaped within a few cycles.
The large change between the first and second cycles is an artificial discontinuity of the model, arising from the reduction of the fracture toughness $G_c^{\mathrm{eff}}$ and the critical energy density $\psi_c^{\mathrm{eff}}$ according to the memory field $S$.
In the analysis below, we therefore exclude the value at the end of the first cycle and quantitatively evaluate the relaxation process from the second cycle onward by exponential fitting.

For the time series $\sigma(c)$, taking the fitting function to be
\begin{equation}
    \sigma(c)=\sigma_{\infty}+A_\sigma\exp\left(-\dfrac{c}{\tau_\theta}\right),
\end{equation}
we estimate the relaxation time $\tau_\theta$ for $c \ge 2$.
We perform the fit by weighted least squares using the error of each cycle.
The relaxation time was found to be $\tau_\theta=3.38\pm 3.4$ cycles.
In the experiments of Goehring et al. \cite{GCA+2010}, which introduced the same quantity, a
maturation via exponential relaxation from an initial transient to a steady-state value has been reported, and the steady-state value after maturation is comparable to that of the present model. The experiment shows about 4 relaxation cycles, which is consistent with the value obtained here.
But the error is slightly large in the present calculation. Thus, we cannot conclude that the simulation shows a perfect reproduction.

\subsection{Evaluation of Energy}

In the previous subsection, we evaluated the geometric maturation using the distribution of crack intersection angles and its standard deviation, and showed that the angle distribution reaches a steady state with a relaxation time of $\tau_\theta=3.38\pm 3.4$ cycles.
In this subsection, we observe the maturation accompanying cycle progression from an energy standpoint.

\begin{figure}
  \begin{subfigure}{0.49\linewidth}
    \centering
    \includegraphics[width=\linewidth]{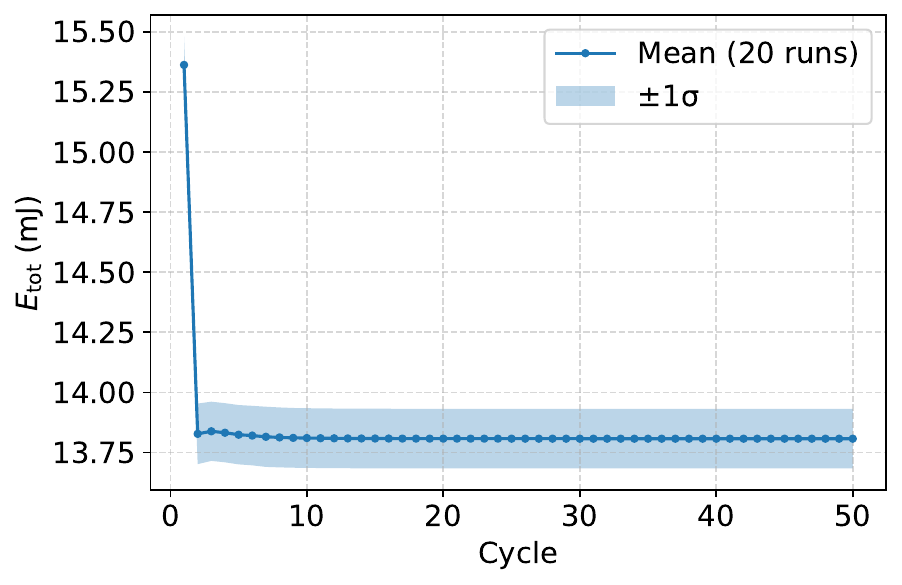}
    \caption{$E_\mathrm{tot}$}
  \end{subfigure}
  \begin{subfigure}{0.49\linewidth}
    \centering
    \includegraphics[width=\linewidth]{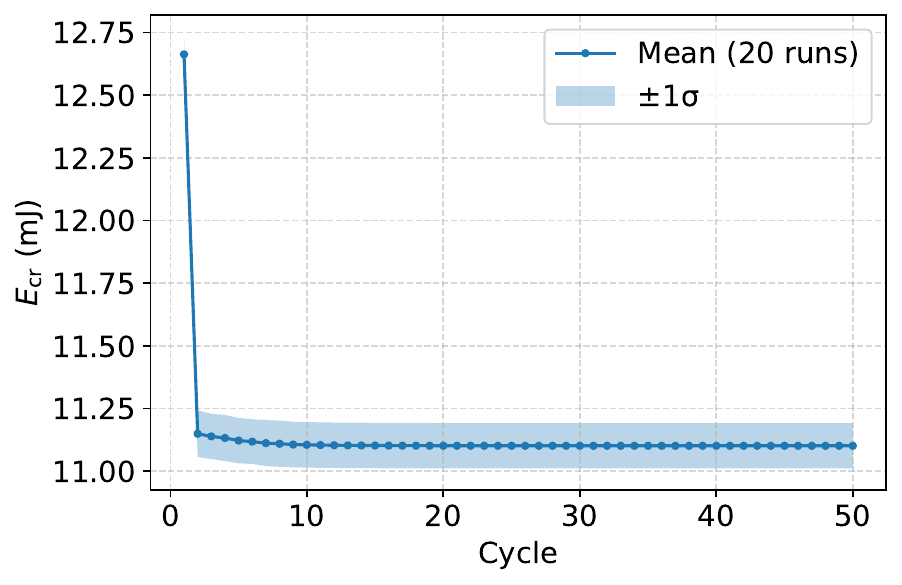}
    \caption{$E_\mathrm{cr}$}
  \end{subfigure}

  \begin{subfigure}{0.49\linewidth}
    \centering
    \includegraphics[width=\linewidth]{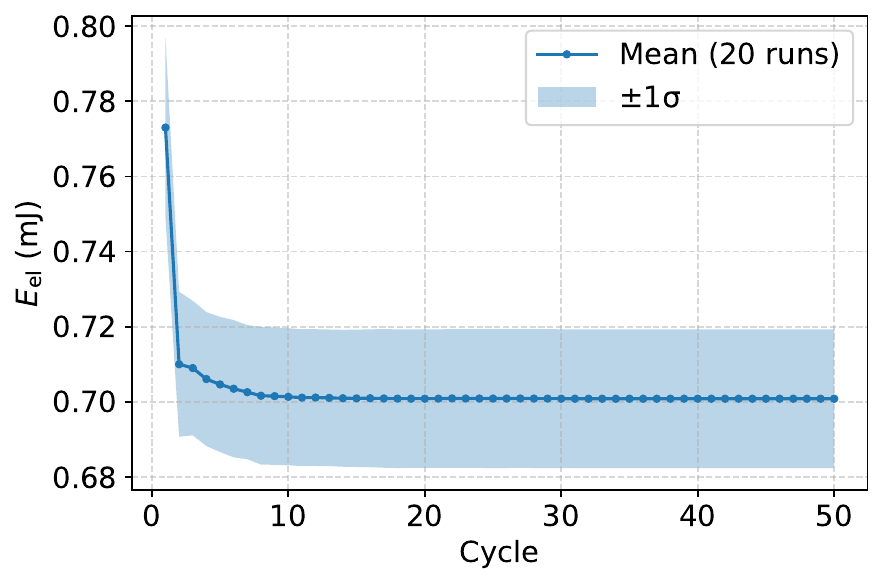}
    \caption{$E_\mathrm{el}$}
  \end{subfigure}
  \begin{subfigure}{0.49\linewidth}
    \centering
    \includegraphics[width=\linewidth]{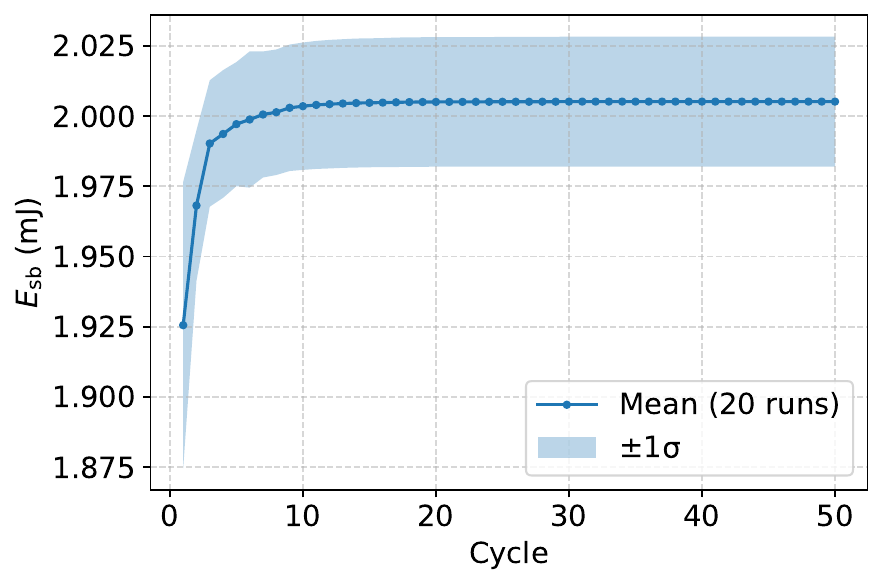}
    \caption{$E_\mathrm{sb}$}
  \end{subfigure}
  \caption{Time series of the individual energy terms evaluated at the end of each cycle, averaged over 20 samples. $E_\mathrm{tot}$ and $E_\mathrm{cr}$ behave very similarly, changing substantially from the first to the second cycle and then gradually approaching a steady-state value. Because in this model the fracture toughness $G_c^{\mathrm{eff}}$ and the critical energy density $\psi_c^{\mathrm{eff}}$ governing fracture in the next cycle are reduced according to the memory field $S$ between cycles, the discontinuous change in $E_\mathrm{tot}$ and $E_{\mathrm{cr}}$ seen in the transition from the first to the second cycle is artificial.
  }
  \label{fig:energy_timeseries}
\end{figure}

First, we define the energies treated in this subsection. We write the total energy of the system as
\begin{equation}
E_\mathrm{tot} = E_\mathrm{el} + E_\mathrm{cr} + E_\mathrm{sb}.
\end{equation}
Each of these terms corresponds to those in Eq.~\eqref{eq:total-energy-functional-for-u}: $E_\mathrm{el}$ is the elastic energy, $E_\mathrm{cr}$ is the crack energy, and $E_\mathrm{sb}$ is the energy from the interaction with the substrate.
At the end of each cycle, we evaluate the energies $E_\mathrm{el}, E_\mathrm{sb}, E_\mathrm{cr}$, as well as the total energy $E_\mathrm{tot}$. Figure~\ref{fig:energy_timeseries} shows the time series of $E_\mathrm{tot}, E_\mathrm{el}, E_\mathrm{sb},$ and $E_\mathrm{cr}$. Here we average over 20 samples and show the standard deviation together with the time series.

The total energy of the system, $E_\mathrm{tot}$, changes substantially from the first to the second cycle and thereafter approaches a steady-state value gradually. The behavior of $E_\mathrm{cr}$ is similar to that of $E_\mathrm{tot}$, showing that the main contribution to $E_\mathrm{tot}$ is the crack-energy term.
The first cycle is a relaxation process dependent on the initial condition, and, as with the time evolution of the angle relaxation, the discontinuous change between the first and second cycles is an artificial model discontinuity.
In the analysis below, we therefore again exclude the value at the end of the first cycle and quantitatively evaluate the relaxation process from the second cycle onward by exponential fitting.

\begin{figure}
  \begin{subfigure}{0.49\linewidth}
    \centering
    \includegraphics[width=\linewidth]{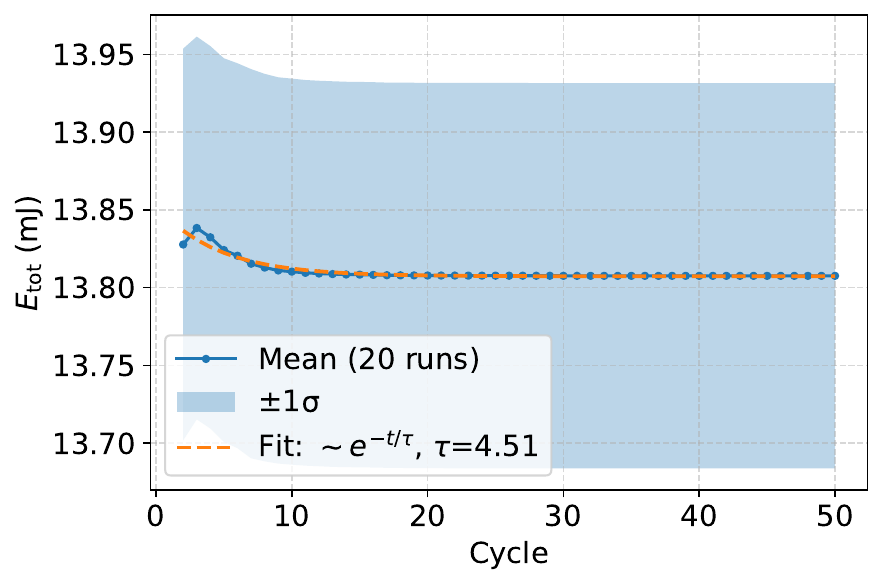}
    \caption{Exponential fitting result for $E_\mathrm{tot}$}
  \end{subfigure}
  \begin{subfigure}{0.49\linewidth}
    \centering
    \includegraphics[width=\linewidth]{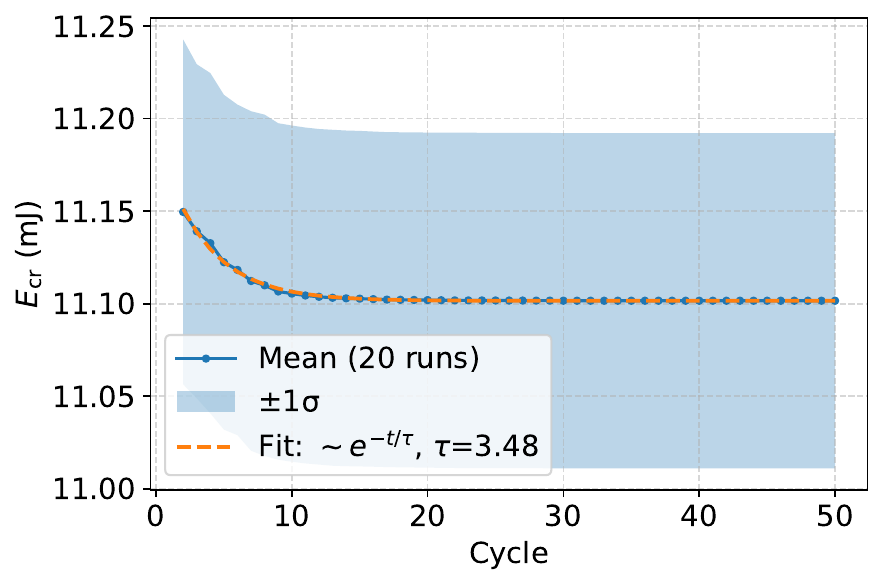}
    \caption{Exponential fitting result for $E_\mathrm{cr}$}
  \end{subfigure}
  \caption{Exponential fitting to the energy change from the second to the fiftieth cycle. (a): $E_\mathrm{tot}$; (b): $E_\mathrm{cr}$. Both $E_\mathrm{tot}$ and $E_\mathrm{cr}$ relax with a relaxation time of a few cycles, and this relaxation time is nearly the same as the relaxation time $\tau_{\theta}$ associated with the angle.}
  \label{fig:energy_expfit}
\end{figure}
Taking the fitting function to be
\begin{equation}
    E(c)=E_\infty + A_E\exp\left(-\dfrac{c}{\tau_E}\right),
\end{equation}
we estimate the relaxation time $\tau_E$ by least squares for $c \ge 2$. The results obtained are plotted together with the data in Fig.~\ref{fig:energy_expfit}. We find that both $E_\mathrm{tot}$ and $E_\mathrm{cr}$ are well fit by an exponential function and relax with a time constant of a few cycles. For the present data, relaxation times
\begin{equation}
  \tau_{E_\mathrm{tot}}= 4.51\pm 0.9, \quad
  \tau_{E_\mathrm{cr}}= 3.48\pm 0.14.
\end{equation}
are obtained. We also find that the relaxation time $\tau_{\theta}$ associated with the angle and the energy-based maturation time scale are essentially equal within error.

\begin{figure}
  \begin{subfigure}{0.49\linewidth}
    \centering
    \includegraphics[width=\linewidth]{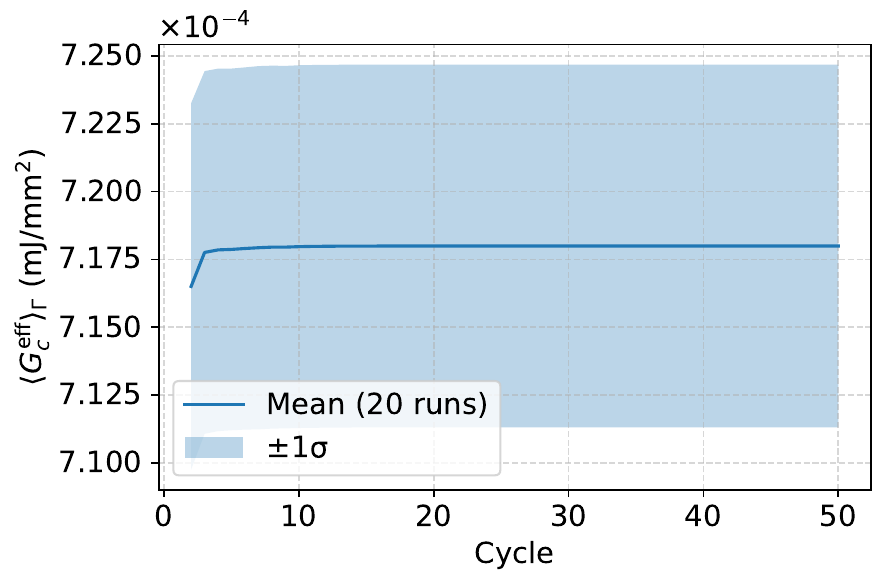}
    \caption{$\ave[\Gamma]{G_c^{\mathrm{eff}}}$ at the end of the cycle}
  \end{subfigure}
  \begin{subfigure}{0.49\linewidth}
    \centering
    \includegraphics[width=\linewidth]{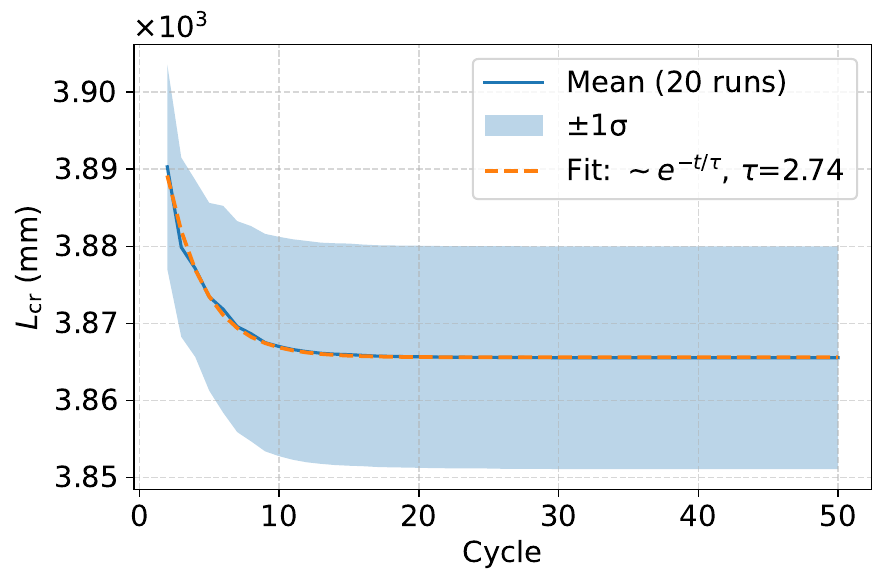}
    \caption{$L_\mathrm{cr}$ at the end of the cycle}
  \end{subfigure}
  \caption{
  (a) The crack-region-weighted average effective fracture toughness $\ave[\Gamma]{G_{c}^{\mathrm{eff}}}$ and (b) the effective crack length $L_{\mathrm{cr}}$, both at the end of the cycle. Each is averaged over 20 samples.
  The effective crack length decays exponentially, and its characteristic relaxation time is slightly smaller than the energy-relaxation timescale.
  }
  \label{fig:relcheck_main_panel}
\end{figure}
We found that the crack energy makes the dominant contribution to the total energy. In this model, the crack energy is defined with the energy functional~\eqref{eq:crack-energy-functional} and the effective fracture toughness~\eqref{eq:effective_gc} as
\begin{equation}
  E_\mathrm{cr}= \int_\Omega \dfrac{3G_c^{\mathrm{eff}}(\mbx)}{8\ell_{0}}
  \left(d(\mbx)+\ell^{2}_{0}\dnorm{\mnab d(\mbx)}^{2}\right)\odif{V}.
\end{equation}
Here we define the effective crack length $L_\mathrm{cr}$ as
\begin{equation}
  L_\mathrm{cr} \equiv \dfrac{3}{8\ell_{0}h}
  \int_\Omega
  \left(d(\mbx)+\ell^{2}_{0}\dnorm{\mnab d(\mbx)}^{2}\right)\odif{V},
\end{equation}
where $h$ is the parameter representing the thickness of the paste. Because the damage field $d$ takes a finite value only at and near the crack, over a width characterized by the length scale $\ell_{0}$, this quantity characterizes the total length of the crack. Using this effective crack length, the crack energy can be decomposed as
\begin{equation}
  E_\mathrm{cr} = \ave[\Gamma]{G_c^{\mathrm{eff}}}hL_{\mathrm{cr}},
  \label{eq:decomp_crackenergy}
\end{equation}
where $\ave[\Gamma]{G_c^{\mathrm{eff}}}$ is the crack-region-weighted average effective fracture toughness,
\begin{equation}
  \ave[\Gamma]{G_c^{\mathrm{eff}}} =
  \dfrac{1}{h L_\mathrm{cr}}
  \int_\Omega \dfrac{3G_c^{\mathrm{eff}}(\mbx)}{8\ell_{0}}
  \left(d(\mbx)+\ell^{2}_{0}\dnorm{\mnab d(\mbx)}^{2}\right)\odif{V}.
\end{equation}
This quantity represents the average value of $G_c^{\mathrm{eff}}$ over the region where the crack has actually formed. A smaller value therefore means the crack has progressed by selecting a path with lower fracture toughness on average.

Using the decomposition of Eq.~\eqref{eq:decomp_crackenergy}, we can discuss whether the change in $E_\mathrm{cr}$ originates from a change in the crack length scale $L_\mathrm{cr}$ or from a change in the effective toughness along the crack, $\langle G_c^{\mathrm{eff}}\rangle_\Gamma$. Figure~\ref{fig:relcheck_main_panel} plots $\langle G_c^{\mathrm{eff}}\rangle_\Gamma$ and $L_\mathrm{cr}$ at the end of each cycle. Here, excluding the artificial change occurring in the transition from the first to the second cycle, we show the results from the second cycle onward.

This result shows that the average effective fracture toughness $\langle G_c^{\mathrm{eff}}\rangle_\Gamma$ is essentially constant, and that changes in the effective crack length $L_{\mathrm{cr}}$ affect the crack energy. The effective crack length also decreases monotonically with the cycle, and fitting this change with an exponential function gives a relaxation time of $2.74\pm 0.14$ cycles. This is slightly smaller than the crack-energy relaxation time.

\begin{figure}
  \centering
  \includegraphics[width=0.80\linewidth]{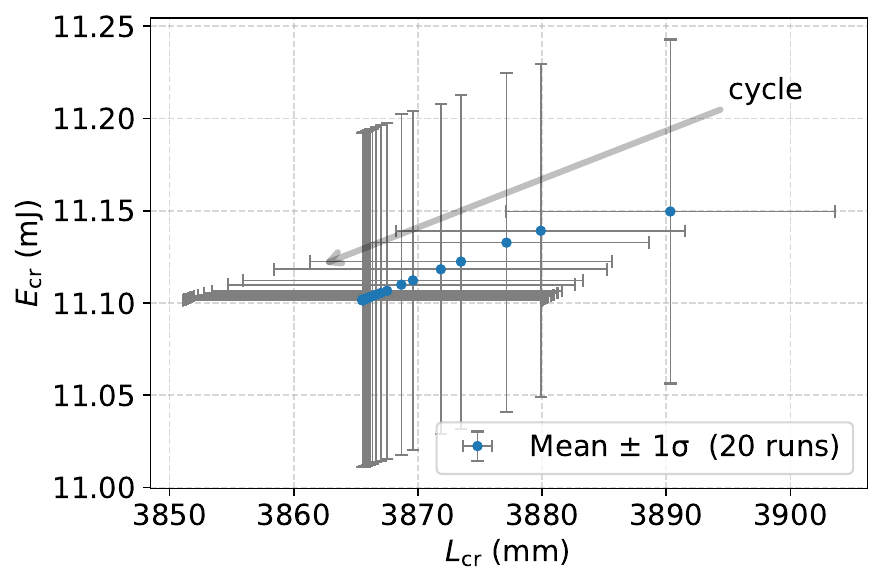}
  \caption{
  Scatter plot of the 20-sample average at the end of each cycle. The horizontal axis is the effective crack length $L_\mathrm{cr}$, and the vertical axis is $E_\mathrm{cr}$. The figure plots data from the second through the fiftieth cycle. As the cycle proceeds, the effective crack length shortens, and the energy value decreases.
  }
  \label{fig:effectiveL_vs_Ecrack}
\end{figure}

From a different perspective, to verify that the change in the crack energy is governed by the effective crack length, we plot $E_\mathrm{cr}$ as a scatter plot against $L_{\mathrm{cr}}$. The result is shown in Fig.~\ref{fig:effectiveL_vs_Ecrack}. The figure shows that, although the trend deviates slightly in the second cycle, from the third cycle onward the effective crack length and the crack energy depend linearly on each other, both decreasing with the cycle and converging toward a certain value.


\section{Summary and Discussion}
\label{sec:summary}

In this paper, building on the phase-field desiccation-fracture model proposed in \cite{HGD2020}, we modeled crack healing during wetting and the effects of previously generated cracks, and performed the corresponding numerical calculations.
We represented crack healing by resetting the relevant physical quantities, and we modeled the scar effect as a local reduction in fracture toughness and critical energy density.
As a result, we found that the initial distribution of crack intersection angles peaks near \ang{90} and shows a broad peak in the range \ang{160} $\sim$ \ang{180}. 
In addition, the frequency near \ang{120} increases as the drying--wetting cycle proceeds.
This increase near \ang{120} is consistent with experimental observations, and we found that as the number of drying--wetting cycles increases, the angle distribution behaves similarly to the experiment.
We further evaluated the standard deviation of the crack angle from \ang{120} and examined it as a function of cycle number; excluding the artificial change introduced by the model, we found a relaxation time of about 3.38 cycles. 
This is the same order as the roughly 4 cycles observed experimentally.
But the error is slightly large in the present calculation. Thus, we need further investigation.

In contrast, compared with the experiment, we also found that the \ang{90} peak in this study does not decrease much. This suggests that a certain fraction of T-shaped intersections persist. In our model, the effective fracture strength for the next cycle is reduced based on the damage field $d$ obtained at the end of the preceding cycle, thereby giving the previous crack path priority as the path for the next crack.
As a result, some of the T-shaped intersections formed early on tend to persist, while the remaining T-shaped intersections relax and transform into Y-shaped intersections.
A quantitative investigation of this behavior would require identifying and tracking the history of individual crack intersections, as well as examining in detail the dependence on the numerical parameters governing the reduction of fracture strength; this has not yet been carried out in the present study and is left for future work.

We also analyzed energy changes. In the model, the total energy decomposes into elastic energy, crack energy, and interaction energy with the substrate. We found that the total energy behavior is dominated mainly by the crack energy. Both the total energy and the crack energy relax exponentially as the cycle proceeds, with nearly equal relaxation times. Moreover, this relaxation time agrees within error with the relaxation time of the standard deviation of the crack angle, indicating that energy relaxation and angle relaxation are essentially the same.

By definition, crack energy equals the product of an effective fracture toughness and an effective crack length. Using this decomposition, we showed that crack-energy relaxation is attributable to the effective crack length shortening as the cycle proceeds.
The shortening of the effective crack length with increasing cycle number appears to be geometrically consistent with the increasing frequency of crack intersection angles near \ang{120}. However, because the effective crack length is defined as an integral of the damage field, it deviates somewhat from, for instance, the apparent crack length seen in the snapshots. For this reason, in order to observe the geometric relaxation of the cracks in the system, it is also necessary to examine changes captured by other indicators, such as the displacement of crack junctions. This paper has not yet examined this and leaves it for future work.

As a further topic for future work, in addition to the points raised above, there remains room to improve the modeling of the scar effect of past cracks. In the present model, no scar exists in the first cycle, and scar-induced weakening is introduced only from the second cycle onward. As a result, the change from the end of the first cycle to the end of the second cycle is artificial, causing the angle variance and energy to behave discontinuously. The change in the angle distribution between the first cycle and subsequent cycles is likewise discontinuous, which does not appear to agree well with the experimental situation. Improving this will require finding more optimal parameters and refining the model itself.


\begin{acknowledgments}

The authors are grateful to Takahiro Hatano, Kazushi Aoyama, Tomohiro Tanogami, and other members of the Hatano lab for valuable discussions. We also thank So Kitsunezaki of Nara Women's University and S.-i. Ito of the University of Tokyo for their advice on this research.
This work was partly supported by JSPS KAKENHI Grant Number 24K06899.
The authors used Claude (Sonnet 5) and ChatGPT (GPT-5) for 
code development and for improving the clarity of the manuscript text. 
The authors reviewed all generated code and revised texts. 
The authors take full responsibility for the content of this paper.

\end{acknowledgments}

\appendix

\section{Definition of Degradation Function and Critical Energy of Crack Nucleation}
\label{sec:degradationFunc}

In this section, we describe the degradation function $g(d)$ from the principle of least action for the phase-field fracture model \cite{LCK2010, PAMM2011, Lor2017, BMMS2014}.
We consider the energy functional that consists of elastic and crack energies to determine the damage field $d(\mbx)$,
\begin{widetext}
\begin{equation}
\Psi[\mbu,d]
= \int_\Omega \left[g(d)H(\mbx,t) + \psi^-(\widetilde{\mbepsilon}(\mbu))\right]\odif{V}
	+\int_\Omega \frac{G_c}{c_0\ell_0}\left(w(d)+\ell_{0}^2 \dnorm{\nabla d}^{2}\right)\odif{V},
\label{eq:E_total_appendix}
\end{equation}
\end{widetext}
where $H(\mbx,t)$ is the history field~\eqref{eq:history-strain-energy}.

To derive the equation that the degradation function must satisfy, we fix the displacement field $\mbu$ and take the variation $\delta d$ with respect to the scalar field $d$. We then derive the Euler--Lagrange equation for the resulting damage field and the condition on the degradation function $g(d)$. 
First, the variation with respect to $d$ of the first term on the right-hand side of Eq.~\eqref{eq:E_total_appendix} (the elastic-energy term) is given by
\begin{equation}
\delta\Psiel^{(d)}[d]
= \int_\Omega g'(d)H(\mbx,t)\delta d\odif{V}.
\end{equation}
For the second term on the right-hand side (the crack-energy term), performing integration by parts on the variation of the gradient term and imposing appropriate boundary conditions on the damage field, we obtain 
\begin{equation}
\delta\Psicr[d] = \int_\Omega
\dfrac{G_c}{c_0\ell_0}
 \left[ w'(d) - 2 \ell_0^2 \Delta d \right]\delta d\odif{V}.
\end{equation}
Thus, the variation of this energy functional with respect to the damage field is
\begin{equation}
\delta\Psi[d]
= \int_\Omega
\left[g'(d)H(\mbx,t) +
\dfrac{G_{c}}{c_{0}\ell_{0}}\left(
w'(d) -{2\ell_0^{2}}\Delta d
\right)
\right]\delta d\odif{V}.
\label{eq:deltaPsi_appendix}
\end{equation}
The necessary and sufficient condition for $\delta\Psi=0$ to hold for arbitrary $\delta d$ is that the quantity inside the brackets in the integrand vanishes, which yields the Euler--Lagrange equation for the damage field $d$,
\begin{equation}
g'(d)H(\mbx,t) + \frac{G_c}{c_0\ell_0}w'(d) - \frac{2G_c\ell_0}{c_0}\Delta d = 0.
\label{eq:EL_d_appendix}
\end{equation}

Next, to evaluate the crack-nucleation threshold, we consider a spatially uniform state. That is, assuming that the crack has scarcely progressed and $d$ is spatially constant, we have $\Delta d = 0$. Linearizing around $d\approx 0$ to describe the infinitesimal damage immediately preceding nucleation, Equation~\eqref{eq:EL_d_appendix} becomes
\begin{equation}
g'(0)H(\mbx,t)+\dfrac{G_c}{c_0\ell_0}w'(0)= 0.
\label{eq:crit_H_general}
\end{equation}
In this situation, $H(\mbx,t)$ is critical at the point where damage begins to grow from the undamaged state, and it equals the critical energy density $\psi_{c}^{\mathrm{eff}}(\mbx)$. 
Since the crack energy adopted in this model is defined with $w(d)=d$, we have $w'(0) = 1$
and $H(\mbx,t) \approx \psi_{c}^{\mathrm{eff}}(\mbx)$, 
we obtain
\begin{equation}
g'(0) \psi_{c}^{\mathrm{eff}}(\mbx)+\dfrac{G_c}{c_0\ell_0}= 0.
\end{equation}
Thus, 
\begin{equation}
g'(0)= -\frac{G_c}{c_0\ell_0\psi_c}
\label{eq:gprime_target}
\end{equation}
follows, where we write $\psi_c^{\mathrm{eff}}(\mbx)=\psi_c$ to simplify.
Because the degradation function reduces the elastic stiffness as $d$ increases, $g'(0)<0$, which is consistent with $\psi_{c}>0$ in this expression.

From the discussion above, the degradation function $g(d)$ for the damage field $d$ is required to satisfy the condition of Eq.~\eqref{eq:gprime_target} within the uniform-field approximation. In the present model, we employ the Lorentz-type degradation function \cite{LCK2010, Lor2017, HGD2020} that satisfies Eq.~\eqref{eq:gprime_target}. 
Then we adopt a degradation function modified so as to ensure the stability of the numerical calculation,
\begin{equation}
g(d)=\dfrac{(1-d)^2}{(1-d)^2+\dfrac{G_c}{c_0\ell_0\psi_c}\,d\,(1+d)}(1-\eta)+\eta.
\label{eq:g_def}
\end{equation}
Here $\eta$ is a small residual parameter introduced to avoid a complete numerical loss of stiffness, and we take $\eta=10^{-6}$. With this choice, the degradation function satisfies $g(0)=1$ in the undamaged state and $g(1)=\eta$ in the fully damaged state. Because of the introduction of the parameter $\eta$, the condition of Eq.~\eqref{eq:gprime_target} is satisfied only approximately, in the form
\begin{equation}
g'(0) = - \dfrac{G_{c}}{c_{0}\ell_{0} \psi_{c}} (1-\eta).
\end{equation}


\section{Procedure for Random Field Generation}
\label{sec:randomfield}

This section details the procedure for generating the two random fields.
Below, we take the following steps:
(1) the paste, described as a two-dimensional elastic body, is discretized on a square lattice with a total of $N$ lattice points, and the coordinates $\mbx_{i,j}$ of the lattice points are fixed;
(2) at each lattice point, mutually correlated random values of the fracture toughness $G_{c}^{(0)}(\mbx_{i,j})$ and the critical energy density $\psi_{c}^{(0)}(\mbx_{i,j})$ are generated independently;
(3) smoothing is used to generate spatially correlated random fields $G_{c}(\mbx_{i,j})$ and $\psi_{c}(\mbx_{i,j})$.

In step (2), the fracture toughness $G_{c}^{(0)}(\mbx_{i,j})$ and critical energy density $\psi_{c}^{(0)}(\mbx_{i,j})$ can be characterized using the mean fracture toughness $\overline{G_{c}^{(0)}}$, the mean critical energy density $\overline{\psi_{c}^{(0)}}$ over the entire paste, the relative standard deviation of the fracture toughness $c_{G_{c}}^{(0)}$, that of the critical energy $c_{\psi_{c}}^{(0)}$, and a parameter $\rho$ representing the correlation between the two random fields.
Here, $\overline{G_{c}^{(0)}}$, $\overline{\psi_{c}^{(0)}}$, $c_{G_{c}}^{(0)}$, $c_{\psi_{c}}^{(0)}$, and $\rho$ are control parameters.

The respective mean quantities are defined by
\begin{align}
\overline{G_c^{(0)}}&=\frac{1}{N}\sum_{i,j}G_c^{(0)}(\mbx_{i,j}), \\
\overline{\psi_c^{(0)}}&=\frac{1}{N}\sum_{i,j}\psi_c^{(0)}(\mbx_{i,j}),
\end{align}
and, writing the respective statistical variances as
\begin{align}
\mathrm{Var}(G_c^{(0)})&=\frac{1}{N}\sum_{i,j}\left(G_c^{(0)}(\mbx_{i,j})-\overline{G_c^{(0)}}\right)^2,\\
\mathrm{Var}(\psi_c^{(0)})&=\frac{1}{N}\sum_{i,j}\left(\psi_c^{(0)}(\mbx_{i,j})-\overline{\psi_c^{(0)}}\right)^2,
\end{align}
the respective relative standard deviations can be expressed as
\begin{equation}
c_{G_c}^{(0)}=\dfrac{\sqrt{\mathrm{Var}(G_c^{(0)})}}{\overline{G_c^{(0)}}},\enspace
c_{\psi_c}^{(0)}=\dfrac{\sqrt{\mathrm{Var}(\psi_c^{(0)})}}{\overline{\psi_c^{(0)}}}.
\end{equation}

In the present model, in order to ensure $G_c>0,\psi_c>0$ at every lattice point, we assume that $G_c^{(0)}(\mbx_{i,j})$ and $\psi_c^{(0)}(\mbx_{i,j})$ follow a log-normal distribution. We describe step (2) below. Since the values are generated independently at each lattice point, we omit the lattice-point coordinate notation $\mbx_{i,j}$.

First, we generate two mutually independent standard normal random numbers $Z_{0}$ and $Z_{1}$ from the standard normal distribution $\mathcal{N}(0,1)$ with mean $0$ and variance $1$. From a linear combination of these, using the parameter $\rho \enspace (0 <  \rho < 1)$, we compute $Z_{2}$ as
\begin{equation}
Z_{2} = \rho Z_{1} + \sqrt{1-\rho^{2}} Z_{0},
\end{equation}
so that $Z_{2}$ is again a standard normal random number with mean $0$ and variance $1$. This linear combination also yields a correlation with covariance $\mathrm{Cov}(Z_{1}, Z_{2}) = \rho$.

Transforming these two quantities $Z_{1}, Z_{2}$ as
\begin{align}
G_{c}^{(0)} & = \exp (\mu_{G_{c}} + \sigma_{G_{c}} Z_{1}) \\
\psi_{c}^{(0)} & = \exp (\mu_{\psi_{c}} + \sigma_{\psi_{c}} Z_{2})
\end{align}
gives $G_{c}^{(0)}, \psi_{c}^{(0)}$ as random numbers that follow a log-normal distribution and are mutually correlated.
Taking
\begin{align}
\sigma_{G_c}& =\sqrt{\ln\left(1+\left(c_{G_c}^{(0)}\right)^{2}\right)},\\
\sigma_{\psi_c}& =\sqrt{\ln\left(1+
\left(c_{\psi_c}^{(0)}\right)^2\right)},\\
\mu_{G_c}& =\ln\overline{G_c^{(0)}}-\frac{1}{2}\sigma_{G_c}^2,\\
\mu_{\psi_c}& =\ln\overline{\psi_c^{(0)}}-\frac{1}{2}\sigma_{\psi_c}^2
\end{align}
gives random numbers for which the means and relative standard deviations of the fracture toughness $G_{c}^{(0)}$ and the critical energy density $\psi_{c}^{(0)}$ are given by $\overline{G_{c}^{(0)}},$ $\overline{\psi_{c}^{(0)}},$ $c_{G_{c}}^{(0)},$ and $c_{\psi_{c}}^{(0)}$, respectively.
In addition, the two random variables have the correlation $\mathrm{Cov}(G_{c}^{(0)},\psi_{c}^{(0)}) = \overline{G_{c}^{(0)}} \overline{\psi_{c}^{(0)}} (e^{\rho \sigma_{G_{c}} \sigma_{\psi_{c}}}-1)$.

Next, as step (3), we impose spatial correlation on the random variables.
The random fields $G_c^{(0)}$ and $\psi_c^{(0)}$ obtained in step (2) are independent at each lattice point and possess no spatial correlation.
We therefore transform them into spatially correlated random fields by smoothing that mimics diffusion. For a random field $f(\mbx_{i,j})$ on a given lattice point, we define the smoothing operation as
\begin{equation}
\cS[f(\mbx_{i,j})]
=\frac{1}{2}f(\mbx_{i,j})+\frac{1}{8}
\Bigl(
f(\mbx_{i+1,j})
+f(\mbx_{i-1,j})
+f(\mbx_{i,j+1})
+f(\mbx_{i,j-1})
\Bigr).
\label{eq:smooth_op}
\end{equation}
Repeating this smoothing operation generates spatial correlation.
In general, the spatial correlation length $\xi$ grows with the number of steps $n$ as $\xi \sim \sqrt{n}$.
This operation also leaves the mean of the random field unchanged, but the variance decreases as the number of repetitions increases.

Writing $\cS^{(N_{s})}$ for the operation of repeating $\cS$ a total of $N_{s}$ times, the final heterogeneous fields $G_c(\mbx),\psi_c(\mbx)$ are given by
\begin{align}
G_{c}(\mbx_{i,j}) & = \cS^{(N_{s})}[G_{c}^{(0)}(\mbx_{i,j})] \\
\psi_{c}(\mbx_{i,j}) & = \cS^{(N_{s})}[\psi_{c}^{(0)}(\mbx_{i,j})].
\end{align}


\section{Details of Numerical Calculation}
\label{sec:details-calc}

In this section, we describe the discretization details for the numerical calculation based on the finite element method (FEM) \cite{Hug2000} used in this study, together with the alternating-iteration algorithm. 

\subsection{Finite Element Method}

\subsubsection{Spatial Discretization}
Dividing the two-dimensional domain $L_x\times L_y$ of the system into $N_x\times N_y$ square elements (four-node bilinear quadrilateral elements of FEM terminology), the lattice sizes $h_x,h_y$ are
\begin{equation}
h_x = \frac{L_x}{N_x},\quad h_y = \frac{L_y}{N_y},
\end{equation}
respectively.
A point on each element is represented using the local coordinates $(\xi,\eta)$ of the parent domain
\begin{equation}
\hat{\Omega}=[-1,1]\times[-1,1].
\end{equation}
Writing the center coordinates of a given square element as $(x_c,y_c)$, the correspondence between the reference (natural) coordinates $(\xi,\eta)$ and the coordinates $(x,y)$ is given by
\begin{equation}
  x = x_c + \frac{h_x}{2}\xi,\quad y = y_c + \frac{h_y}{2}\eta.
  \label{eq:mapping_xi_xy}
\end{equation}

We take the shape functions $N_a(\xi,\eta),\enspace (a=1,\dots,4)$ on the four-node bilinear quadrilateral element to be
\begin{align}
  N_1(\xi,\eta) &= \dfrac{1}{4}(1-\xi)(1-\eta), \nonumber\\
  N_2(\xi,\eta) &= \dfrac{1}{4}(1+\xi)(1-\eta), \nonumber\\
  N_3(\xi,\eta) &= \dfrac{1}{4}(1+\xi)(1+\eta), \nonumber\\
  N_4(\xi,\eta) &= \dfrac{1}{4}(1-\xi)(1+\eta).
  \label{eq:shape-functions}
\end{align}

In this model, the values of the displacement field $\mbu=(u_x,u_y)$, the damage field $d$, the two random fields $G_c$ and $\psi_c$ representing the heterogeneity of the paste, and the memory field $S$ are held at the four nodes $a\enspace(a=1,\dots,4; (\xi_a,\eta_a)=(\pm 1,\pm 1))$ of each bilinear quadrilateral element.
The history field $H$, on the other hand, is held at the four Gauss quadrature nodes $p\enspace (p=1,\dots,4;(\xi_p,\eta_p)=(\pm 1/\sqrt{3},\pm 1/\sqrt{3}) )$ of each bilinear quadrilateral element.

In the finite element approximation, fields are interpolated using their values at the nodes. For instance, the displacement field $\mbu=(u_x,u_y)$ at the coordinates $\mbx=(x,y)$ is interpolated, using the values $\mbu_a^e=(u_{x,a}^e,u_{y,a}^e)$ at the corresponding element $e$ and its nodes $a$, as
\begin{equation}
  \mbu(\mbx) \approx \sum_{a=1}^4 N_a(\xi,\eta)\mbu_a^e.
\end{equation}
Similarly, the damage field $d$, the random fields $G_c,\psi_c$, and the memory field $S$ are interpolated as
\begin{align}
    d(\mbx)      &\approx \sum_{a=1}^4 N_a(\xi,\eta)d_a^e, \\
    G_c(\mbx)    &\approx \sum_{a=1}^4 N_a(\xi,\eta)G_{c,a}^e, \\
    \psi_c(\mbx) &\approx \sum_{a=1}^4 N_a(\xi,\eta)\psi_{c,a}^e,\\
    S(\mbx)      &\approx \sum_{a=1}^4 N_a(\xi,\eta)S_a^e,
\end{align}
respectively.
Derivatives of the field quantities are interpolated using derivatives of the shape functions.

\subsubsection{Gaussian Quadrature Formula}

In this study, the volume integral over each element is approximated by a $2\times2$ Gaussian quadrature formula. We take the Gauss quadrature nodes $p$ on element $e$ to be $(\xi_{p},\eta_{p}) = (\pm 1/\sqrt{3},\pm 1/\sqrt{3})\enspace  (p=1,\dots,4)$, with weight $w_p=1$, and use the Jacobian $h_xh_y/4$.

Hereafter, for an arbitrary function $f(\xi,\eta)$ on element $e$, we write
\begin{equation}
    f\vert_{p} \equiv f(\xi_p,\eta_p)
\end{equation}
to denote the value of $f$ evaluated at the Gauss node $(\xi_p,\eta_p)$.
For fields whose values are held at the nodes, interpolation uses the shape functions, while for fields whose values are held at the Gauss nodes, those values are used directly.
When a derivative of a field is required, it is evaluated using the derivatives of the shape functions, as described above. With this, the integral over element $e$ is given by the Gaussian quadrature formula as
\begin{equation}
  \int_{\Omega_e} f \odif{V} \approx \sum_{p=1}^{4} f\vert_p w_p
  \dfrac{h_{x}h_{y}}{4}  h
  = \sum_{p=1}^{4} f\vert_p w_v.
\end{equation}
Here $h$ is the thickness of the paste, and
\begin{equation}
  w_v = \dfrac{h_{x}h_{y}h}{4},
\end{equation}
because $w_p = 1$.

\subsubsection{Weak Form}

The displacement field and damage field that minimize the total energy functional satisfy
\begin{equation}
  \fdv{\Psitot[\mbu,d]}{\mbu} = 0, \enspace
  \fdv{\Psitot^{(d)}[\mbu,d]}{d} = 0,
\end{equation}
from the energy functionals~\eqref{eq:total-energy-functional-for-u} and \eqref{eq:total-energy-functional-for-d}.
In the finite element method, introducing the weighting functions $\mbW(\mbx)=(W_{x}(\mbx),W_{y}(\mbx)),W(\mbx) $, these two equations are solved in weak form as
\begin{widetext}
\begin{align}
0 & = \int \fdv{\Psitot}{u_{x}} W_{x} \odif{V}\\
  & = \int \left[
    \left(
      g(d)K\ave[+]{\tr(\widetilde{\mbepsilon})}
      +K\ave[-]{\tr(\widetilde{\mbepsilon})}
      +2g(d) \mu e_{xx}
    \right) \partial_x W_x
    +2 g(d) \mu e_{xy}\partial_y W_x
    +\dfrac{\kappa}{h}g(d) u_x W_x \right]\odif{V},\\
  \label{eq:weak_ux}
0 & = \int \fdv{\Psitot}{u_{y}} W_{y} \odif{V}\\
  & = \int \left[
    \left(
      g(d)K\ave[+]{\tr(\widetilde{\mbepsilon})}
      +K \ave[-]{\tr(\widetilde{\mbepsilon})}
      +2g(d) \mu e_{yy}
    \right) \partial_y W_y
    + 2 g(d) \mu e_{xy}\partial_x W_y
    + \dfrac{\kappa}{h}g(d) u_y W_y \right]\odif{V},\\
  \label{eq:weak_uy}
\text{and}\quad
0 & = \int \fdv{\Psitot^{(d)}}{d} W \odif{V}\\
  & = \int \left[
    g'(d) H W + \dfrac{G_c^{\mathrm{eff}}}{c_0\ell_0}W
    +\dfrac{2G_c^{\mathrm{eff}}\ell_0}{c_0} \nabla d \cdot \nabla W
    + \dfrac{\kappa}{2h} g'(d) \dnorm{\mbu}^2W
  \right] \odif{V},
\end{align}
\end{widetext}
where $e_{xx},e_{yy},e_{xy}$ are the $xx,yy,xy$ components of the shearing strain $\mbe$.

We interpolate the displacement and damage fields with the shape functions
and express the strain tensor in terms of the shape functions.
In addition, the integral is approximated using the Gaussian quadrature formula.
Finally, taking the shape functions as the weighting functions, the weak-form integrals (residual vector) for the displacement field at node $a$ of element $e$ for the displacement field are
\begin{widetext}
\begin{multline}
  R_{a}^{(u_x),e}(\{\mbu_{a}^{e},d_{a}^{e}\}) =
  \sum_{p=1}^{4}\Big[
    \left(
      g(d)K\ave[+]{\tr(\widetilde{\mbepsilon})}
      + K \ave[-]{\tr(\widetilde{\mbepsilon})}
      +2 g(d)\mu e_{xx}
    \right) \partial_x  N_a\\
    + 2 g(d) \mu e_{xy} \partial_y N_a
    + \dfrac{\kappa}{h} g(d) u_x  N_a
    \Big]\Bigg\vert_{p} w_v,
\end{multline}
and
\begin{multline}
  R_{a}^{(u_y),e}(\{\mbu_{a}^{e},d_{a}^{e}\}) =
  \sum_{p=1}^{4}\Big[
    \left(
      g(d)K\ave[+]{\tr(\widetilde{\mbepsilon})}
      + K \ave[-]{\tr(\widetilde{\mbepsilon})}
      +2g(d)\mu e_{yy}
    \right) \partial_y  N_a\\
    + 2 g(d) \mu e_{xy} \partial_x N_a
    + \dfrac{\kappa}{h} g(d) u_y  N_a
    \Big]\Bigg\vert_{p} w_v
\end{multline}
for each component.
The weak-form integral (residual vector) at node $a$ of element $e$ for the damage field is likewise
\begin{equation}
  R_{a}^{(d),e}(\{\mbu_{a}^{e},d_{a}^{e}\})=\sum_{p=1}^{4}
  \Big[ g'(d)H N_a
    + \dfrac{G^{\mathrm{eff}}_{c}}{c_0\ell_0} N_a
    + \dfrac{2G^{\mathrm{eff}}_{c}\ell_0}{c_0}\nabla d\cdot\nabla N_a
    + \dfrac{\kappa}{2h}\dnorm{\mbu}^2  g'(d)N_a\Big]\Bigg\vert_{p} w_v.
\end{equation}
\end{widetext}
Collecting these over nodes, we obtain the residual vectors $\mbR^{(\mbu)}$ with $2N_xN_y$ dimension, and $\mbR^{(d)}$ with $N_x N_y$ dimension.

To solve numerically, we also need the Jacobian (tangent stiffness matrix).
For the displacement field for element $e$,
we differentiate the residual for the $i$-component of the displacement field at node $a$ with respect to the $j$-component of the displacement field at node $b$ and we obtain
\begin{widetext}
\begin{multline}
K_{ab}^{(u_i,u_j),e} = \pdv{R_{a}^{(u_i),e}}{u_{j,b}^{e}}
= \sum_{p=1}^{4} \Big[
  [g(d)\Theta(\tr(\widetilde{\varepsilon})) + \Theta(-\tr(\widetilde{\varepsilon})) ]
  K \partial_i N_a \partial_j N_b + g(d)\mu (2\delta_{ij}-1) \partial_i N_a \partial_j N_b
\\
+ g(d) \mu (\partial_{\bar{\imath}} N_a)(\partial_{\bar{\jmath}} N_b)
+\dfrac{\kappa}{h} g(d) \delta_{ij} N_a N_b
\Big]\Bigg\vert_{p} w_v.
\end{multline}
\end{widetext}
Collecting this over nodes and directions, we construct the matrix $\mbK^{(\mbu)}$, which is a $2N_xN_y$ by $2N_xN_y$ square matrix.
Here $\Theta(x)$ is the step function, and $\bar{\imath}$ is an index such that $\bar{\imath}=y$ when $i=x$ and $\bar{\imath}=x$ when $i=y$; $\bar{\jmath}$ is defined analogously.

For the damage field, we obtain
\begin{widetext}
\[
K_{ab}^{(d),e} = \pdv{R^{(d),e}_{a}}{d_{b}^{e}}
=
\sum_{p=1}^{4}
  \Big[ g''(d)H N_a N_b
    + \dfrac{2G^{\mathrm{eff}}_{c}\ell_0}{c_0}\nabla N_b \cdot\nabla N_a
    + \dfrac{\kappa}{2h}\dnorm{\mbu}^2  g''(d)N_aN_b\Big]\Bigg\vert_{p} w_v.
\]
\end{widetext}
Collecting this over nodes, we construct the matrix $\mbK^{(d)}$, which is a $N_xN_y$ by $N_xN_y$ square matrix.

\subsection{Alternating Iteration Scheme at Each Step}

In this study, we introduce discrete steps $n=0,1,2,\dots, N_\text{step}$ into the desiccation process, and at each step change the shrinkage strain $\varepsilon_\text{sh}$ by a small amount. Writing the maximum number of steps as $N_\text{step}$ and the absolute value of the maximum shrinkage strain as $\varepsilon_\text{sh}^\text{max}$, the shrinkage strain $\varepsilon_\text{sh}^{n}$ at step $n$ is given by
\begin{equation}
  \varepsilon_\text{sh}^{n} =
  - \dfrac{n}{N_\text{step}} \varepsilon_\text{sh}^\text{max}.
\end{equation}
At each step, we first specify the corresponding shrinkage strain and solve the coupled nonlinear problem for the displacement and damage fields using an alternating iteration method.
In this section, we denote the solution vectors as $\vec{u}$
for the displacement field with dimension $2N_xN_y$ and $\vec{d}$ for the damage field with dimension $N_xN_y$.

In the crack calculation from step $n$ to $n+1$, we perform the alternating iteration.
At the beginning of the iteration, we take the initial values as the converged solution $\vec{u}^n, \vec{d}^n$ from the previous step and the history field $\vec{H}^{n}$ obtained from it,
\begin{equation}
  \vec{u}^{(0)}=\vec{u}^n,\quad \vec{d}^{(0)}=\vec{d}^n,\quad \vec{H}^{(0)}=\vec{H}^n.
\end{equation}

The alternating iteration process with the iteration count $k$ is as follows:
\begin{enumerate}
  \item \textbf{$u$ subproblem (damage field $d$ fixed)}:
  With the damage field $\vec{d}^{(k)}$ fixed, $\vec{u}^{(k+1)}$ is obtained using the residual vector $\mbR^{(\mbu)}$ and the tangent stiffness matrix $\mbK^{(\mbu)}$.
  In this study, the nonlinear system of equations for $u$ is solved using the Newton method with PCG and backtracking line search\cite{NW2006}.
  \item \textbf{Update of the history field $H$}:
  $\psiel^{+}$ is evaluated at each Gauss point $\mbx_p$ from the obtained $\vec{u}^{(k+1)}$, and 	the update
  \begin{equation}
    H^{(k+1)}(\mbx_p)  =\max\left(H^{(k)}(\mbx_p),\,\psiel^{+}(\mbx_p)\right)
  \end{equation}
  is performed.
  \item \textbf{$d$ subproblem (displacement field and history field $\mbu,H$ fixed)}:
  With $\vec{u}^{(k+1)}$ and $\vec{H}^{(k+1)}$ fixed, $d$ is solved using the residual vector $\mbR^{(d)}$ and the tangent stiffness matrix $\mbK^{(d)}$, under the constraint $d^{(k)}(\mbx_{i,j}) \le d^{(k+1)}(\mbx_{i,j}) \le 1$ at each point, to obtain $\vec{d}^{(k+1)}$.
  This constraint prevents the crack from recovering due to numerical instabilities and thereby guarantees the irreversibility of crack generation. 
  This was solved using the projected Newton method with PCG. 
  \item \textbf{Convergence criterion}:
  For
  \begin{equation}
    \Delta\vec{u}=\vec{u}^{(k+1)}-\vec{u}^{(k)},\quad
    \Delta \vec{d}=\vec{d}^{(k+1)}-\vec{d}^{(k)},
  \end{equation}
  we obtain the maximum values of the updates to the displacement and the damage as
  \begin{equation}
    \Delta u_\text{max}=\max_i\norm{\Delta u_i},\quad
    \Delta d_\text{max}=\max_i\norm{\Delta d_i}.
  \end{equation}
  If
  \begin{equation}
    \Delta u_{\max}< \epsilon_\text{tol}^{u}, \quad \Delta d_{\max}<\epsilon_\text{tol}^{d}
  \end{equation}
  is satisfied, the iteration has converged.
  In this implementation, we take $\epsilon_\text{tol}^{u}=10^{-6}$ and $\epsilon_\text{tol}^{d}=10^{-4}$.
 If convergence has not been reached, the procedure continues with $k\to k+1$.
\end{enumerate}
Finally, to guarantee the irreversibility of crack generation even after the step update, the constraint
\begin{equation}
   \max(d^{n+1}(\mbx_{i,j}), d^{n}(\mbx_{i,j})) = d^{n+1}(\mbx_{i,j})
\end{equation}
was imposed at each point.


\section{Angle Measurement Algorithm}
\label{sec:angle-algorithm}

In this section, we describe the method for measuring the crack intersection angle.
We treat a snapshot of the damage field $d$ as a $1024 \times 1024$ pixel image, which is sufficiently fine compared with the discretized lattice $128 \times 128$.
We apply a Gaussian filter to blur the crack image and then perform Guo--Hall thinning \cite{GH1992} using the OpenCV library \cite{opencv_library} until the cracks are down to a width of a single pixel.
After that, to detect crack intersections, which we call nodes, we extract pixels in the thinned image whose 8-neighborhood contains three or more crack pixels.

Using the resulting set of nodes together with the thinned image, we trace along the thinned crack from each node to its neighboring node to extract the pixels of the thin-line segment, which we call an edge, connecting the two junctions.
Then the edge $i$ that originated from node $n$ can be represented by the pixel sequence
\begin{equation}
    P_i = (p_0, p_1, p_2,\dots),
\end{equation}
where the first point $p_0$ corresponds to node $n$.

To measure the angle, it is necessary to express the direction in which a crack extends from a node as a vector. However, pixels near a node are densely clustered. These pixels make an error. Therefore, we construct a representative outward vector without pixels near the node as follows.

For each node $n$, we define a set of pixels representing its core region.
Specifically, we construct the core region by the thin-line pixels within a disk of radius three pixels centered on the node.
We also add pixels to the core region if they have three or more crack pixels in the surrounding 8-neighborhood and exist in the disk.
We remove the core region pixels from the pixel sequence $P_i$.
From the remaining point sequence $\widetilde{P_i}=(p_j,p_{j+1},\dots) \enspace (j> 0)$,
we construct a short line segment $(q_0,q_1,\dots,q_{m-1})$, where
$q_0 = p_{j+r}$ and $q_{m-1} = p_{j+r+m-1}$. $r$ and $m$ are numerical parameters taken to be $r=2, m=9$, respectively.
Next, we calculate the average direction $\mbt_i$ from the short line segments as
\begin{equation}
\mbt_i = \dfrac{\mbq_{m-1}-\mbq_{0}}{\dnorm{\mbq_{m-1}-\mbq_{0}}},
\end{equation}
where $\mbq_j$ is a coordinate vector of the pixel $q_{j}$.
Here we use this $\mbt_i$ as a direction of the crack represented by the sequence $P_i$.

The intersection angle $\theta_{ij}$ between edge $i$ and $j$ is straightforwardly determined by
\begin{equation}
\theta_{ij} = \cos^{-1}
\left(
{\mathbf{t}_i \cdot \mathbf{t}_j}
\right),
\quad (\ang{0} \le \theta_{ij} \le \ang{180}).
\end{equation}

\bibliography{bibs}

\begin{thebibliography}{43}%
\makeatletter
\providecommand \@ifxundefined [1]{%
 \@ifx{#1\undefined}
}%
\providecommand \@ifnum [1]{%
 \ifnum #1\expandafter \@firstoftwo
 \else \expandafter \@secondoftwo
 \fi
}%
\providecommand \@ifx [1]{%
 \ifx #1\expandafter \@firstoftwo
 \else \expandafter \@secondoftwo
 \fi
}%
\providecommand \natexlab [1]{#1}%
\providecommand \enquote  [1]{``#1''}%
\providecommand \bibnamefont  [1]{#1}%
\providecommand \bibfnamefont [1]{#1}%
\providecommand \citenamefont [1]{#1}%
\providecommand \href@noop [0]{\@secondoftwo}%
\providecommand \href [0]{\begingroup \@sanitize@url \@href}%
\providecommand \@href[1]{\@@startlink{#1}\@@href}%
\providecommand \@@href[1]{\endgroup#1\@@endlink}%
\providecommand \@sanitize@url [0]{\catcode `\\12\catcode `\$12\catcode
  `\&12\catcode `\#12\catcode `\^12\catcode `\_12\catcode `\%12\relax}%
\providecommand \@@startlink[1]{}%
\providecommand \@@endlink[0]{}%
\providecommand \url  [0]{\begingroup\@sanitize@url \@url }%
\providecommand \@url [1]{\endgroup\@href {#1}{\urlprefix }}%
\providecommand \urlprefix  [0]{URL }%
\providecommand \Eprint [0]{\href }%
\providecommand \doibase [0]{https://doi.org/}%
\providecommand \selectlanguage [0]{\@gobble}%
\providecommand \bibinfo  [0]{\@secondoftwo}%
\providecommand \bibfield  [0]{\@secondoftwo}%
\providecommand \translation [1]{[#1]}%
\providecommand \BibitemOpen [0]{}%
\providecommand \bibitemStop [0]{}%
\providecommand \bibitemNoStop [0]{.\EOS\space}%
\providecommand \EOS [0]{\spacefactor3000\relax}%
\providecommand \BibitemShut  [1]{\csname bibitem#1\endcsname}%
\let\auto@bib@innerbib\@empty
\bibitem [{\citenamefont {Goehring}\ \emph {et~al.}(2015)\citenamefont
  {Goehring}, \citenamefont {Nakahara}, \citenamefont {Dutta}, \citenamefont
  {Kitsunezaki},\ and\ \citenamefont {Tarafdar}}]{GND+2015}%
  \BibitemOpen
  \bibfield  {author} {\bibinfo {author} {\bibfnamefont {L.}~\bibnamefont
  {Goehring}}, \bibinfo {author} {\bibfnamefont {A.}~\bibnamefont {Nakahara}},
  \bibinfo {author} {\bibfnamefont {T.}~\bibnamefont {Dutta}}, \bibinfo
  {author} {\bibfnamefont {S.}~\bibnamefont {Kitsunezaki}},\ and\ \bibinfo
  {author} {\bibfnamefont {S.}~\bibnamefont {Tarafdar}},\ }\href@noop {} {\emph
  {\bibinfo {title} {Desiccation Cracks and Their Patterns: Formation and
  Modelling in Science and Nature}}},\ Statistical Physics of Fracture and
  Breakdown\ (\bibinfo  {publisher} {Wiley-VCH, Verlag GmbH \& Co. KGaA},\
  \bibinfo {address} {Weinheim},\ \bibinfo {year} {2015})\BibitemShut {NoStop}%
\bibitem [{\citenamefont {Corte}\ and\ \citenamefont {Higashi}(1964)}]{CH1964}%
  \BibitemOpen
  \bibfield  {author} {\bibinfo {author} {\bibfnamefont {A.~E.}\ \bibnamefont
  {Corte}}\ and\ \bibinfo {author} {\bibfnamefont {A.}~\bibnamefont
  {Higashi}},\ }\href
  {https://usace.contentdm.oclc.org/digital/collection/p266001coll1/id/6153/}
  {\emph {\bibinfo {title} {Experimental research on desiccation cracks in
  soil}}},\ \bibinfo {type} {Tech. Rep.}\ \bibinfo {number} {66}\ (\bibinfo
  {institution} {U.S. Army Materiel Command, Cold Regions Research {\&}
  Engineering Laboratory},\ \bibinfo {address} {Hanover, New Hampshire},\
  \bibinfo {year} {1964})\BibitemShut {NoStop}%
\bibitem [{\citenamefont {Groisman}\ and\ \citenamefont
  {Kaplan}(1994)}]{GK1994}%
  \BibitemOpen
  \bibfield  {author} {\bibinfo {author} {\bibfnamefont {A.}~\bibnamefont
  {Groisman}}\ and\ \bibinfo {author} {\bibfnamefont {E.}~\bibnamefont
  {Kaplan}},\ }\href {https://doi.org/10.1209/0295-5075/25/6/004} {\bibfield
  {journal} {\bibinfo  {journal} {Europhysics Letters (EPL)}\ }\textbf
  {\bibinfo {volume} {25}},\ \bibinfo {pages} {415} (\bibinfo {year}
  {1994})}\BibitemShut {NoStop}%
\bibitem [{\citenamefont {Allain}\ and\ \citenamefont {Limat}(1995)}]{AL1995}%
  \BibitemOpen
  \bibfield  {author} {\bibinfo {author} {\bibfnamefont {C.}~\bibnamefont
  {Allain}}\ and\ \bibinfo {author} {\bibfnamefont {L.}~\bibnamefont {Limat}},\
  }\href {https://doi.org/10.1103/PhysRevLett.74.2981} {\bibfield  {journal}
  {\bibinfo  {journal} {Phys. Rev. Lett.}\ }\textbf {\bibinfo {volume} {74}},\
  \bibinfo {pages} {2981} (\bibinfo {year} {1995})}\BibitemShut {NoStop}%
\bibitem [{\citenamefont {Shorlin}\ \emph {et~al.}(2000)\citenamefont
  {Shorlin}, \citenamefont {de~Bruyn}, \citenamefont {Graham},\ and\
  \citenamefont {Morris}}]{SdBGM2000}%
  \BibitemOpen
  \bibfield  {author} {\bibinfo {author} {\bibfnamefont {K.~A.}\ \bibnamefont
  {Shorlin}}, \bibinfo {author} {\bibfnamefont {J.~R.}\ \bibnamefont
  {de~Bruyn}}, \bibinfo {author} {\bibfnamefont {M.}~\bibnamefont {Graham}},\
  and\ \bibinfo {author} {\bibfnamefont {S.~W.}\ \bibnamefont {Morris}},\
  }\href {https://doi.org/10.1103/PhysRevE.61.6950} {\bibfield  {journal}
  {\bibinfo  {journal} {Phys. Rev. E}\ }\textbf {\bibinfo {volume} {61}},\
  \bibinfo {pages} {6950} (\bibinfo {year} {2000})}\BibitemShut {NoStop}%
\bibitem [{\citenamefont {Bohn}\ \emph {et~al.}(2005)\citenamefont {Bohn},
  \citenamefont {Pauchard},\ and\ \citenamefont {Couder}}]{BPC2005}%
  \BibitemOpen
  \bibfield  {author} {\bibinfo {author} {\bibfnamefont {S.}~\bibnamefont
  {Bohn}}, \bibinfo {author} {\bibfnamefont {L.}~\bibnamefont {Pauchard}},\
  and\ \bibinfo {author} {\bibfnamefont {Y.}~\bibnamefont {Couder}},\ }\href
  {https://doi.org/10.1103/PhysRevE.71.046214} {\bibfield  {journal} {\bibinfo
  {journal} {Physical Review E}\ }\textbf {\bibinfo {volume} {71}},\ \bibinfo
  {pages} {046214} (\bibinfo {year} {2005})}\BibitemShut {NoStop}%
\bibitem [{\citenamefont {Pasricha}\ \emph {et~al.}(2009)\citenamefont
  {Pasricha}, \citenamefont {Wad}, \citenamefont {Pasricha},\ and\
  \citenamefont {Ogale}}]{PWPO2009}%
  \BibitemOpen
  \bibfield  {author} {\bibinfo {author} {\bibfnamefont {K.}~\bibnamefont
  {Pasricha}}, \bibinfo {author} {\bibfnamefont {U.}~\bibnamefont {Wad}},
  \bibinfo {author} {\bibfnamefont {R.}~\bibnamefont {Pasricha}},\ and\
  \bibinfo {author} {\bibfnamefont {S.}~\bibnamefont {Ogale}},\ }\href
  {https://doi.org/10.1016/j.physa.2008.12.039} {\bibfield  {journal} {\bibinfo
   {journal} {Physica A: Statistical Mechanics and its Applications}\ }\textbf
  {\bibinfo {volume} {388}},\ \bibinfo {pages} {1352} (\bibinfo {year}
  {2009})}\BibitemShut {NoStop}%
\bibitem [{\citenamefont {Nakahara}\ and\ \citenamefont
  {Matsuo}(2005)}]{NM2005}%
  \BibitemOpen
  \bibfield  {author} {\bibinfo {author} {\bibfnamefont {A.}~\bibnamefont
  {Nakahara}}\ and\ \bibinfo {author} {\bibfnamefont {Y.}~\bibnamefont
  {Matsuo}},\ }\href {https://doi.org/10.1143/JPSJ.74.1362} {\bibfield
  {journal} {\bibinfo  {journal} {Journal of the Physical Society of Japan}\
  }\textbf {\bibinfo {volume} {74}},\ \bibinfo {pages} {1362} (\bibinfo {year}
  {2005})}\BibitemShut {NoStop}%
\bibitem [{\citenamefont {Nakahara}\ and\ \citenamefont
  {Matsuo}(2006{\natexlab{a}})}]{NM2006}%
  \BibitemOpen
  \bibfield  {author} {\bibinfo {author} {\bibfnamefont {A.}~\bibnamefont
  {Nakahara}}\ and\ \bibinfo {author} {\bibfnamefont {Y.}~\bibnamefont
  {Matsuo}},\ }\href {https://doi.org/10.1103/PhysRevE.74.045102} {\bibfield
  {journal} {\bibinfo  {journal} {Physical Review E}\ }\textbf {\bibinfo
  {volume} {74}},\ \bibinfo {pages} {045102} (\bibinfo {year}
  {2006}{\natexlab{a}})}\BibitemShut {NoStop}%
\bibitem [{\citenamefont {Nakahara}\ and\ \citenamefont
  {Matsuo}(2006{\natexlab{b}})}]{NM2006a}%
  \BibitemOpen
  \bibfield  {author} {\bibinfo {author} {\bibfnamefont {A.}~\bibnamefont
  {Nakahara}}\ and\ \bibinfo {author} {\bibfnamefont {Y.}~\bibnamefont
  {Matsuo}},\ }\href {https://doi.org/10.1088/1742-5468/2006/07/P07016}
  {\bibfield  {journal} {\bibinfo  {journal} {Journal of Statistical Mechanics:
  Theory and Experiment}\ }\textbf {\bibinfo {volume} {2006}},\ \bibinfo
  {pages} {P07016} (\bibinfo {year} {2006}{\natexlab{b}})}\BibitemShut
  {NoStop}%
\bibitem [{\citenamefont {M\"uller}(1998)}]{Mul1998a}%
  \BibitemOpen
  \bibfield  {author} {\bibinfo {author} {\bibfnamefont {G.}~\bibnamefont
  {M\"uller}},\ }\href
  {https://doi.org/https://doi.org/10.1016/S0377-0273(98)00045-6} {\bibfield
  {journal} {\bibinfo  {journal} {Journal of Volcanology and Geothermal
  Research}\ }\textbf {\bibinfo {volume} {86}},\ \bibinfo {pages} {93}
  (\bibinfo {year} {1998})}\BibitemShut {NoStop}%
\bibitem [{\citenamefont {M{\"u}ller}(1998)}]{Mul1998}%
  \BibitemOpen
  \bibfield  {author} {\bibinfo {author} {\bibfnamefont {G.}~\bibnamefont
  {M{\"u}ller}},\ }\href {https://doi.org/10.1029/98JB00389} {\bibfield
  {journal} {\bibinfo  {journal} {Journal of Geophysical Research: Solid
  Earth}\ }\textbf {\bibinfo {volume} {103}},\ \bibinfo {pages} {15239}
  (\bibinfo {year} {1998})}\BibitemShut {NoStop}%
\bibitem [{\citenamefont {Toramaru}\ and\ \citenamefont
  {Matsumoto}(2004)}]{TM2004a}%
  \BibitemOpen
  \bibfield  {author} {\bibinfo {author} {\bibfnamefont {A.}~\bibnamefont
  {Toramaru}}\ and\ \bibinfo {author} {\bibfnamefont {T.}~\bibnamefont
  {Matsumoto}},\ }\href {https://doi.org/10.1029/2003JB002686} {\bibfield
  {journal} {\bibinfo  {journal} {Journal of Geophysical Research: Solid
  Earth}\ }\textbf {\bibinfo {volume} {109}},\ \bibinfo {pages} {B02205}
  (\bibinfo {year} {2004})}\BibitemShut {NoStop}%
\bibitem [{\citenamefont {Goehring}\ \emph {et~al.}(2006)\citenamefont
  {Goehring}, \citenamefont {Morris},\ and\ \citenamefont {Lin}}]{GML2006}%
  \BibitemOpen
  \bibfield  {author} {\bibinfo {author} {\bibfnamefont {L.}~\bibnamefont
  {Goehring}}, \bibinfo {author} {\bibfnamefont {S.~W.}\ \bibnamefont
  {Morris}},\ and\ \bibinfo {author} {\bibfnamefont {Z.}~\bibnamefont {Lin}},\
  }\href {https://doi.org/10.1103/PhysRevE.74.036115} {\bibfield  {journal}
  {\bibinfo  {journal} {Physical Review E}\ }\textbf {\bibinfo {volume} {74}},\
  \bibinfo {pages} {036115} (\bibinfo {year} {2006})}\BibitemShut {NoStop}%
\bibitem [{\citenamefont {Weinberger}(1999)}]{Wei1999}%
  \BibitemOpen
  \bibfield  {author} {\bibinfo {author} {\bibfnamefont {R.}~\bibnamefont
  {Weinberger}},\ }\href
  {https://doi.org/https://doi.org/10.1016/S0191-8141(99)00029-2} {\bibfield
  {journal} {\bibinfo  {journal} {Journal of Structural Geology}\ }\textbf
  {\bibinfo {volume} {21}},\ \bibinfo {pages} {379} (\bibinfo {year}
  {1999})}\BibitemShut {NoStop}%
\bibitem [{\citenamefont {Goehring}\ \emph {et~al.}(2009)\citenamefont
  {Goehring}, \citenamefont {Mahadevan},\ and\ \citenamefont
  {Morris}}]{GMM2009}%
  \BibitemOpen
  \bibfield  {author} {\bibinfo {author} {\bibfnamefont {L.}~\bibnamefont
  {Goehring}}, \bibinfo {author} {\bibfnamefont {L.}~\bibnamefont
  {Mahadevan}},\ and\ \bibinfo {author} {\bibfnamefont {S.~W.}\ \bibnamefont
  {Morris}},\ }\href {https://doi.org/10.1073/pnas.0805132106} {\bibfield
  {journal} {\bibinfo  {journal} {Proceedings of the National Academy of
  Sciences}\ }\textbf {\bibinfo {volume} {106}},\ \bibinfo {pages} {387}
  (\bibinfo {year} {2009})}\BibitemShut {NoStop}%
\bibitem [{\citenamefont {Goehring}\ \emph {et~al.}(2010)\citenamefont
  {Goehring}, \citenamefont {Conroy}, \citenamefont {Akhter}, \citenamefont
  {Clegg},\ and\ \citenamefont {Routh}}]{GCA+2010}%
  \BibitemOpen
  \bibfield  {author} {\bibinfo {author} {\bibfnamefont {L.}~\bibnamefont
  {Goehring}}, \bibinfo {author} {\bibfnamefont {R.}~\bibnamefont {Conroy}},
  \bibinfo {author} {\bibfnamefont {A.}~\bibnamefont {Akhter}}, \bibinfo
  {author} {\bibfnamefont {W.~J.}\ \bibnamefont {Clegg}},\ and\ \bibinfo
  {author} {\bibfnamefont {A.~F.}\ \bibnamefont {Routh}},\ }\href
  {https://doi.org/10.1039/b922206e} {\bibfield  {journal} {\bibinfo  {journal}
  {Soft Matter}\ }\textbf {\bibinfo {volume} {6}},\ \bibinfo {pages} {3562}
  (\bibinfo {year} {2010})}\BibitemShut {NoStop}%
\bibitem [{\citenamefont {Goehring}(2013)}]{Goe2013}%
  \BibitemOpen
  \bibfield  {author} {\bibinfo {author} {\bibfnamefont {L.}~\bibnamefont
  {Goehring}},\ }\href {https://doi.org/10.1098/rsta.2012.0353} {\bibfield
  {journal} {\bibinfo  {journal} {Philosophical Transactions of the Royal
  Society A: Mathematical, Physical and Engineering Sciences}\ }\textbf
  {\bibinfo {volume} {371}},\ \bibinfo {pages} {20120353} (\bibinfo {year}
  {2013})}\BibitemShut {NoStop}%
\bibitem [{\citenamefont {Kitsunezaki}(1999)}]{Kit1999}%
  \BibitemOpen
  \bibfield  {author} {\bibinfo {author} {\bibfnamefont {S.}~\bibnamefont
  {Kitsunezaki}},\ }\href {https://doi.org/10.1103/PhysRevE.60.6449} {\bibfield
   {journal} {\bibinfo  {journal} {Physical Review E}\ }\textbf {\bibinfo
  {volume} {60}},\ \bibinfo {pages} {6449} (\bibinfo {year}
  {1999})}\BibitemShut {NoStop}%
\bibitem [{\citenamefont {Jagla}(2004)}]{Jag2004}%
  \BibitemOpen
  \bibfield  {author} {\bibinfo {author} {\bibfnamefont {E.~A.}\ \bibnamefont
  {Jagla}},\ }\href {https://doi.org/10.1103/PhysRevE.69.056212} {\bibfield
  {journal} {\bibinfo  {journal} {Physical Review E}\ }\textbf {\bibinfo
  {volume} {69}},\ \bibinfo {pages} {056212} (\bibinfo {year}
  {2004})}\BibitemShut {NoStop}%
\bibitem [{\citenamefont {Ito}\ and\ \citenamefont
  {Yukawa}(2014{\natexlab{a}})}]{IY2014}%
  \BibitemOpen
  \bibfield  {author} {\bibinfo {author} {\bibfnamefont {S.-i.}\ \bibnamefont
  {Ito}}\ and\ \bibinfo {author} {\bibfnamefont {S.}~\bibnamefont {Yukawa}},\
  }\href {https://doi.org/10.1103/PhysRevE.90.042909} {\bibfield  {journal}
  {\bibinfo  {journal} {Physical Review E}\ }\textbf {\bibinfo {volume} {90}},\
  \bibinfo {pages} {042909} (\bibinfo {year} {2014}{\natexlab{a}})}\BibitemShut
  {NoStop}%
\bibitem [{\citenamefont {Hu}\ \emph {et~al.}(2020)\citenamefont {Hu},
  \citenamefont {Guilleminot},\ and\ \citenamefont {Dolbow}}]{HGD2020}%
  \BibitemOpen
  \bibfield  {author} {\bibinfo {author} {\bibfnamefont {T.}~\bibnamefont
  {Hu}}, \bibinfo {author} {\bibfnamefont {J.}~\bibnamefont {Guilleminot}},\
  and\ \bibinfo {author} {\bibfnamefont {J.~E.}\ \bibnamefont {Dolbow}},\
  }\href {https://doi.org/10.1016/j.cma.2020.113106} {\bibfield  {journal}
  {\bibinfo  {journal} {Computer Methods in Applied Mechanics and Engineering}\
  }\textbf {\bibinfo {volume} {368}},\ \bibinfo {pages} {113106} (\bibinfo
  {year} {2020})}\BibitemShut {NoStop}%
\bibitem [{\citenamefont {Hornig}\ \emph {et~al.}(1996)\citenamefont {Hornig},
  \citenamefont {Sokolov},\ and\ \citenamefont {Blumen}}]{HSB1996}%
  \BibitemOpen
  \bibfield  {author} {\bibinfo {author} {\bibfnamefont {T.}~\bibnamefont
  {Hornig}}, \bibinfo {author} {\bibfnamefont {I.~M.}\ \bibnamefont
  {Sokolov}},\ and\ \bibinfo {author} {\bibfnamefont {A.}~\bibnamefont
  {Blumen}},\ }\href {https://doi.org/10.1103/PhysRevE.54.4293} {\bibfield
  {journal} {\bibinfo  {journal} {Physical Review E}\ }\textbf {\bibinfo
  {volume} {54}},\ \bibinfo {pages} {4293} (\bibinfo {year}
  {1996})}\BibitemShut {NoStop}%
\bibitem [{\citenamefont {Hal{\'a}sz}\ \emph {et~al.}(2017)\citenamefont
  {Hal{\'a}sz}, \citenamefont {Nakahara}, \citenamefont {Kitsunezaki},\ and\
  \citenamefont {Kun}}]{HNKK2017}%
  \BibitemOpen
  \bibfield  {author} {\bibinfo {author} {\bibfnamefont {Z.}~\bibnamefont
  {Hal{\'a}sz}}, \bibinfo {author} {\bibfnamefont {A.}~\bibnamefont
  {Nakahara}}, \bibinfo {author} {\bibfnamefont {S.}~\bibnamefont
  {Kitsunezaki}},\ and\ \bibinfo {author} {\bibfnamefont {F.}~\bibnamefont
  {Kun}},\ }\href {https://doi.org/10.1103/PhysRevE.96.033006} {\bibfield
  {journal} {\bibinfo  {journal} {Physical Review E}\ }\textbf {\bibinfo
  {volume} {96}},\ \bibinfo {pages} {033006} (\bibinfo {year}
  {2017})}\BibitemShut {NoStop}%
\bibitem [{\citenamefont {Haque}\ \emph {et~al.}(2023)\citenamefont {Haque},
  \citenamefont {Mitra}, \citenamefont {Tarafdar},\ and\ \citenamefont
  {Dutta}}]{HMTD2023}%
  \BibitemOpen
  \bibfield  {author} {\bibinfo {author} {\bibfnamefont {R.~A.~I.}\
  \bibnamefont {Haque}}, \bibinfo {author} {\bibfnamefont {A.~J.}\ \bibnamefont
  {Mitra}}, \bibinfo {author} {\bibfnamefont {S.}~\bibnamefont {Tarafdar}},\
  and\ \bibinfo {author} {\bibfnamefont {T.}~\bibnamefont {Dutta}},\ }\href
  {https://doi.org/10.1016/j.chaos.2023.113894} {\bibfield  {journal} {\bibinfo
   {journal} {Chaos, Solitons \& Fractals}\ }\textbf {\bibinfo {volume}
  {174}},\ \bibinfo {pages} {113894} (\bibinfo {year} {2023})},\ \Eprint
  {https://arxiv.org/abs/2305.01991} {arXiv:2305.01991 [cond-mat]} \BibitemShut
  {NoStop}%
\bibitem [{\citenamefont {Szatm{\'a}ri}\ \emph {et~al.}(2025)\citenamefont
  {Szatm{\'a}ri}, \citenamefont {Nakahara}, \citenamefont {Kitsunezaki},\ and\
  \citenamefont {Kun}}]{SNKK2025}%
  \BibitemOpen
  \bibfield  {author} {\bibinfo {author} {\bibfnamefont {R.}~\bibnamefont
  {Szatm{\'a}ri}}, \bibinfo {author} {\bibfnamefont {A.}~\bibnamefont
  {Nakahara}}, \bibinfo {author} {\bibfnamefont {S.}~\bibnamefont
  {Kitsunezaki}},\ and\ \bibinfo {author} {\bibfnamefont {F.}~\bibnamefont
  {Kun}},\ }\href {https://doi.org/10.21468/SciPostPhys.19.6.142} {\bibfield
  {journal} {\bibinfo  {journal} {SciPost Physics}\ }\textbf {\bibinfo {volume}
  {19}},\ \bibinfo {pages} {142} (\bibinfo {year} {2025})}\BibitemShut
  {NoStop}%
\bibitem [{\citenamefont {Ito}\ and\ \citenamefont
  {Yukawa}(2014{\natexlab{b}})}]{IY2014a}%
  \BibitemOpen
  \bibfield  {author} {\bibinfo {author} {\bibfnamefont {S.-i.}\ \bibnamefont
  {Ito}}\ and\ \bibinfo {author} {\bibfnamefont {S.}~\bibnamefont {Yukawa}},\
  }\href {https://doi.org/10.7566/JPSJ.83.124005} {\bibfield  {journal}
  {\bibinfo  {journal} {Journal of the Physical Society of Japan}\ }\textbf
  {\bibinfo {volume} {83}},\ \bibinfo {pages} {124005} (\bibinfo {year}
  {2014}{\natexlab{b}})}\BibitemShut {NoStop}%
\bibitem [{\citenamefont {Hofmann}\ \emph {et~al.}(2015)\citenamefont
  {Hofmann}, \citenamefont {Anderssohn}, \citenamefont {Bahr}, \citenamefont
  {Wei{\ss}},\ and\ \citenamefont {Nellesen}}]{HAB+2015}%
  \BibitemOpen
  \bibfield  {author} {\bibinfo {author} {\bibfnamefont {M.}~\bibnamefont
  {Hofmann}}, \bibinfo {author} {\bibfnamefont {R.}~\bibnamefont {Anderssohn}},
  \bibinfo {author} {\bibfnamefont {H.-A.}\ \bibnamefont {Bahr}}, \bibinfo
  {author} {\bibfnamefont {H.-J.}\ \bibnamefont {Wei{\ss}}},\ and\ \bibinfo
  {author} {\bibfnamefont {J.}~\bibnamefont {Nellesen}},\ }\href
  {https://doi.org/10.1103/PhysRevLett.115.154301} {\bibfield  {journal}
  {\bibinfo  {journal} {Physical Review Letters}\ }\textbf {\bibinfo {volume}
  {115}},\ \bibinfo {pages} {154301} (\bibinfo {year} {2015})}\BibitemShut
  {NoStop}%
\bibitem [{\citenamefont {Amor}\ \emph {et~al.}(2009)\citenamefont {Amor},
  \citenamefont {Marigo},\ and\ \citenamefont {Maurini}}]{AMM2009}%
  \BibitemOpen
  \bibfield  {author} {\bibinfo {author} {\bibfnamefont {H.}~\bibnamefont
  {Amor}}, \bibinfo {author} {\bibfnamefont {J.-J.}\ \bibnamefont {Marigo}},\
  and\ \bibinfo {author} {\bibfnamefont {C.}~\bibnamefont {Maurini}},\ }\href
  {https://doi.org/10.1016/j.jmps.2009.04.011} {\bibfield  {journal} {\bibinfo
  {journal} {Journal of the Mechanics and Physics of Solids}\ }\textbf
  {\bibinfo {volume} {57}},\ \bibinfo {pages} {1209} (\bibinfo {year}
  {2009})}\BibitemShut {NoStop}%
\bibitem [{\citenamefont {Miehe}\ \emph {et~al.}(2010)\citenamefont {Miehe},
  \citenamefont {Hofacker},\ and\ \citenamefont {Welschinger}}]{MHW2010}%
  \BibitemOpen
  \bibfield  {author} {\bibinfo {author} {\bibfnamefont {C.}~\bibnamefont
  {Miehe}}, \bibinfo {author} {\bibfnamefont {M.}~\bibnamefont {Hofacker}},\
  and\ \bibinfo {author} {\bibfnamefont {F.}~\bibnamefont {Welschinger}},\
  }\href {https://doi.org/10.1016/j.cma.2010.04.011} {\bibfield  {journal}
  {\bibinfo  {journal} {Computer Methods in Applied Mechanics and Engineering}\
  }\textbf {\bibinfo {volume} {199}},\ \bibinfo {pages} {2765} (\bibinfo {year}
  {2010})}\BibitemShut {NoStop}%
\bibitem [{\citenamefont {Ambrosio}\ and\ \citenamefont
  {Tortorelli}(1990)}]{AT1990}%
  \BibitemOpen
  \bibfield  {author} {\bibinfo {author} {\bibfnamefont {L.}~\bibnamefont
  {Ambrosio}}\ and\ \bibinfo {author} {\bibfnamefont {V.~M.}\ \bibnamefont
  {Tortorelli}},\ }\href {https://doi.org/10.1002/cpa.3160430805} {\bibfield
  {journal} {\bibinfo  {journal} {Communications on Pure and Applied
  Mathematics}\ }\textbf {\bibinfo {volume} {43}},\ \bibinfo {pages} {999}
  (\bibinfo {year} {1990})}\BibitemShut {NoStop}%
\bibitem [{\citenamefont {Bourdin}\ \emph {et~al.}(2000)\citenamefont
  {Bourdin}, \citenamefont {Francfort},\ and\ \citenamefont
  {Marigo}}]{BFM2000}%
  \BibitemOpen
  \bibfield  {author} {\bibinfo {author} {\bibfnamefont {B.}~\bibnamefont
  {Bourdin}}, \bibinfo {author} {\bibfnamefont {G.}~\bibnamefont {Francfort}},\
  and\ \bibinfo {author} {\bibfnamefont {J.-J.}\ \bibnamefont {Marigo}},\
  }\href {https://doi.org/10.1016/S0022-5096(99)00028-9} {\bibfield  {journal}
  {\bibinfo  {journal} {Journal of the Mechanics and Physics of Solids}\
  }\textbf {\bibinfo {volume} {48}},\ \bibinfo {pages} {797} (\bibinfo {year}
  {2000})}\BibitemShut {NoStop}%
\bibitem [{\citenamefont {Griffith}(1921)}]{Gri1921}%
  \BibitemOpen
  \bibfield  {author} {\bibinfo {author} {\bibfnamefont {A.~A.}\ \bibnamefont
  {Griffith}},\ }\href {https://doi.org/10.1098/rsta.1921.0006} {\bibfield
  {journal} {\bibinfo  {journal} {Philosophical Transactions of the Royal
  Society of London. Series A, Containing Papers of a Mathematical or Physical
  Character}\ }\textbf {\bibinfo {volume} {221}},\ \bibinfo {pages} {163}
  (\bibinfo {year} {1921})}\BibitemShut {NoStop}%
\bibitem [{\citenamefont {Pham}\ \emph {et~al.}(2011)\citenamefont {Pham},
  \citenamefont {Amor}, \citenamefont {Marigo},\ and\ \citenamefont
  {Maurini}}]{PAMM2011}%
  \BibitemOpen
  \bibfield  {author} {\bibinfo {author} {\bibfnamefont {K.}~\bibnamefont
  {Pham}}, \bibinfo {author} {\bibfnamefont {H.}~\bibnamefont {Amor}}, \bibinfo
  {author} {\bibfnamefont {J.-J.}\ \bibnamefont {Marigo}},\ and\ \bibinfo
  {author} {\bibfnamefont {C.}~\bibnamefont {Maurini}},\ }\href
  {https://doi.org/10.1177/1056789510386852} {\bibfield  {journal} {\bibinfo
  {journal} {International Journal of Damage Mechanics}\ }\textbf {\bibinfo
  {volume} {20}},\ \bibinfo {pages} {618} (\bibinfo {year} {2011})}\BibitemShut
  {NoStop}%
\bibitem [{\citenamefont {Bourdin}\ \emph {et~al.}(2014)\citenamefont
  {Bourdin}, \citenamefont {Marigo}, \citenamefont {Maurini},\ and\
  \citenamefont {Sicsic}}]{BMMS2014}%
  \BibitemOpen
  \bibfield  {author} {\bibinfo {author} {\bibfnamefont {B.}~\bibnamefont
  {Bourdin}}, \bibinfo {author} {\bibfnamefont {J.-J.}\ \bibnamefont {Marigo}},
  \bibinfo {author} {\bibfnamefont {C.}~\bibnamefont {Maurini}},\ and\ \bibinfo
  {author} {\bibfnamefont {P.}~\bibnamefont {Sicsic}},\ }\href
  {https://doi.org/10.1103/PhysRevLett.112.014301} {\bibfield  {journal}
  {\bibinfo  {journal} {Physical Review Letters}\ }\textbf {\bibinfo {volume}
  {112}},\ \bibinfo {pages} {014301} (\bibinfo {year} {2014})}\BibitemShut
  {NoStop}%
\bibitem [{\citenamefont {Liang}\ \emph {et~al.}(2003)\citenamefont {Liang},
  \citenamefont {Huang}, \citenamefont {Prévost},\ and\ \citenamefont
  {Suo}}]{LHPS2003}%
  \BibitemOpen
  \bibfield  {author} {\bibinfo {author} {\bibfnamefont {J.}~\bibnamefont
  {Liang}}, \bibinfo {author} {\bibfnamefont {R.}~\bibnamefont {Huang}},
  \bibinfo {author} {\bibfnamefont {J.}~\bibnamefont {Prévost}},\ and\
  \bibinfo {author} {\bibfnamefont {Z.}~\bibnamefont {Suo}},\ }\href
  {https://doi.org/10.1016/S0020-7683(03)00095-7} {\bibfield  {journal}
  {\bibinfo  {journal} {International Journal of Solids and Structures}\
  }\textbf {\bibinfo {volume} {40}},\ \bibinfo {pages} {2343} (\bibinfo {year}
  {2003})}\BibitemShut {NoStop}%
\bibitem [{\citenamefont {Ahrens}\ \emph {et~al.}(2005)\citenamefont {Ahrens},
  \citenamefont {Geveci},\ and\ \citenamefont {Law}}]{ParaView}%
  \BibitemOpen
  \bibfield  {author} {\bibinfo {author} {\bibfnamefont {J.}~\bibnamefont
  {Ahrens}}, \bibinfo {author} {\bibfnamefont {B.}~\bibnamefont {Geveci}},\
  and\ \bibinfo {author} {\bibfnamefont {C.}~\bibnamefont {Law}},\ }\bibinfo
  {title} {Visualization handbook}\ (\bibinfo  {publisher} {Elsevier Inc.},\
  \bibinfo {address} {Burlington, MA, USA},\ \bibinfo {year} {2005})\ Chap.\
  \bibinfo {chapter} {{ParaView}: An End-User Tool for Large Data
  Visualization}, pp.\ \bibinfo {pages} {717--731}\BibitemShut {NoStop}%
\bibitem [{\citenamefont {Lorentz}\ \emph {et~al.}(2011)\citenamefont
  {Lorentz}, \citenamefont {Cuvilliez},\ and\ \citenamefont
  {Kazymyrenko}}]{LCK2010}%
  \BibitemOpen
  \bibfield  {author} {\bibinfo {author} {\bibfnamefont {E.}~\bibnamefont
  {Lorentz}}, \bibinfo {author} {\bibfnamefont {S.}~\bibnamefont {Cuvilliez}},\
  and\ \bibinfo {author} {\bibfnamefont {K.}~\bibnamefont {Kazymyrenko}},\
  }\href {https://doi.org/10.1016/j.crme.2010.10.010} {\bibfield  {journal}
  {\bibinfo  {journal} {Comptes Rendus. Mécanique}\ }\textbf {\bibinfo
  {volume} {339}},\ \bibinfo {pages} {20} (\bibinfo {year} {2011})}\BibitemShut
  {NoStop}%
\bibitem [{\citenamefont {Lorentz}(2017)}]{Lor2017}%
  \BibitemOpen
  \bibfield  {author} {\bibinfo {author} {\bibfnamefont {E.}~\bibnamefont
  {Lorentz}},\ }\href {https://doi.org/10.1007/s10704-017-0225-z} {\bibfield
  {journal} {\bibinfo  {journal} {International Journal of Fracture}\ }\textbf
  {\bibinfo {volume} {207}},\ \bibinfo {pages} {123} (\bibinfo {year}
  {2017})}\BibitemShut {NoStop}%
\bibitem [{\citenamefont {Hughes}(2000)}]{Hug2000}%
  \BibitemOpen
  \bibfield  {author} {\bibinfo {author} {\bibfnamefont {T.~J.~R.}\
  \bibnamefont {Hughes}},\ }\href@noop {} {\emph {\bibinfo {title} {The Finite
  Element Method: Linear Static and Dynamic Finite Element Analysis}}}\
  (\bibinfo  {publisher} {Dover Publications},\ \bibinfo {address} {Mineola,
  NY},\ \bibinfo {year} {2000})\BibitemShut {NoStop}%
\bibitem [{\citenamefont {Nocedal}\ and\ \citenamefont
  {Wright}(2006)}]{NW2006}%
  \BibitemOpen
  \bibfield  {author} {\bibinfo {author} {\bibfnamefont {J.}~\bibnamefont
  {Nocedal}}\ and\ \bibinfo {author} {\bibfnamefont {S.~J.}\ \bibnamefont
  {Wright}},\ }\href@noop {} {\emph {\bibinfo {title} {Numerical
  Optimization}}},\ \bibinfo {edition} {2nd}\ ed.,\ Springer Series in
  Operations Research\ (\bibinfo  {publisher} {Springer},\ \bibinfo {address}
  {New York},\ \bibinfo {year} {2006})\BibitemShut {NoStop}%
\bibitem [{\citenamefont {Guo}\ and\ \citenamefont {Hall}(1992)}]{GH1992}%
  \BibitemOpen
  \bibfield  {author} {\bibinfo {author} {\bibfnamefont {Z.}~\bibnamefont
  {Guo}}\ and\ \bibinfo {author} {\bibfnamefont {R.~W.}\ \bibnamefont {Hall}},\
  }\href {https://doi.org/10.1016/1049-9660(92)90029-3} {\bibfield  {journal}
  {\bibinfo  {journal} {CVGIP: Image Understanding}\ }\textbf {\bibinfo
  {volume} {55}},\ \bibinfo {pages} {317} (\bibinfo {year} {1992})}\BibitemShut
  {NoStop}%
\bibitem [{\citenamefont {Bradski}(2000)}]{opencv_library}%
  \BibitemOpen
  \bibfield  {author} {\bibinfo {author} {\bibfnamefont {G.}~\bibnamefont
  {Bradski}},\ }\href@noop {} {\bibfield  {journal} {\bibinfo  {journal} {Dr.
  Dobb's Journal of Software Tools}\ } (\bibinfo {year} {2000})}\BibitemShut
  {NoStop}%
\end{thebibliography}%
\end{document}